\documentclass[runningheads]{llncs}
\usepackage{eccv}
\usepackage{eccvabbrv}
\usepackage{graphicx}
\usepackage{booktabs}
\usepackage[accsupp]{axessibility}
\usepackage{hyperref}
\usepackage{orcidlink}
\usepackage{pifont}
\usepackage{multirow} 
\usepackage{url}
\usepackage{booktabs}
\usepackage{amsfonts}
\usepackage{nicefrac}
\usepackage{microtype}
\usepackage{multirow}
\usepackage{graphicx}
\usepackage{wrapfig}
\usepackage{enumitem}
\usepackage[table]{xcolor}
\usepackage{colortbl}
\usepackage{subcaption}
\usepackage{mathtools}
\usepackage{bm}
\renewcommand{\paragraph}[1]{\smallskip\noindent\textbf{#1.}~} %
\usepackage{xcolor}
\usepackage{tcolorbox}
\definecolor{promptpurple}{HTML}{7A1E8A}
\definecolor{promptblue}{HTML}{7EC3EA}
\definecolor{promptblue2}{HTML}{E7F3FF}
\newcounter{promptcount}
\renewcommand{\thepromptcount}{\arabic{promptcount}}

\newtcolorbox{promptbox}[2][]{%
  float=t,
  colframe=promptblue2,
  colback=white,
  colbacktitle=promptblue2,
  coltitle=black, 
  step=promptcount, 
  title=\textbf{Prompt \thepromptcount: #2},
  arc=4pt,
  boxrule=3pt,
  left=6pt,
  right=6pt,
  top=6pt,
  bottom=6pt,
  fonttitle=\bfseries,
  label=#1
}

\usepackage[linesnumbered,ruled,vlined]{algorithm2e}
\usepackage{algpseudocode}

\SetKwComment{Comment}{$\triangleright$\ }{}
\begin{document}

\title{MG$^2$-RAG: Multi-Granularity Graph for Multimodal Retrieval-Augmented Generation} 
\titlerunning{MG$^2$-RAG: Multi-Granularity Graph for MM-RAG}
\author{
Sijun~Dai\inst{1,}\textsuperscript{\dag}\orcidlink{0009-0003-7912-0987} \and
Qiang~Huang\inst{1,}\textsuperscript{\dag}\orcidlink{0000-0003-1120-4685} \and
Xiaoxing~You\inst{2}\orcidlink{0009-0002-3700-6759} \and
Jun~Yu\inst{1,}\textsuperscript{*}\orcidlink{0009-0005-2316-5478}
}
\authorrunning{Sijun Dai et al.}
\institute{School of Intelligence Science and Engineering,\\
Harbin Institute of Technology (Shenzhen)\\
\email{daisijun@stu.hit.edu.cn, \{huangqiang, yujun\}@hit.edu.cn} \and
School of Computer Science, Hangzhou Dianzi University\\
\email{youxiaoxing@hdu.edu.cn} \\
\textsuperscript{\dag}Equal contribution \quad
\textsuperscript{*}Corresponding author
}
\maketitle
\begin{abstract}
Retrieval-Augmented Generation (RAG) mitigates hallucinations in Multimodal Large Language Models (MLLMs), yet existing systems struggle with complex cross-modal reasoning.
Flat vector retrieval often ignores structural dependencies, while current graph-based methods rely on costly ``translation-to-text'' pipelines that discard fine-grained visual information.
To address these limitations, we propose \textbf{MG$^2$-RAG}, a lightweight \textbf{M}ulti-\textbf{G}ranularity \textbf{G}raph \textbf{RAG} framework that jointly improves graph construction, modality fusion, and cross-modal retrieval.
MG$^2$-RAG constructs a hierarchical multimodal knowledge graph by combining lightweight textual parsing with entity-driven visual grounding, enabling textual entities and visual regions to be fused into unified multimodal nodes that preserve atomic evidence.
Building on this representation, we introduce a multi-granularity graph retrieval mechanism that aggregates dense similarities and propagates relevance across the graph to support structured multi-hop reasoning.
Extensive experiments across four representative multimodal tasks (i.e., retrieval, knowledge-based VQA, reasoning, and classification) demonstrate that MG$^2$-RAG consistently achieves state-of-the-art performance while reducing graph construction overhead with an average 43.3$\times$ speedup and 23.9$\times$ cost reduction compared with advanced graph-based frameworks.
The source code is publicly available at \url{https://github.com/Daboolu/MG2-RAG}.

\end{abstract}

\section{Introduction}
\label{sec:intro}

Multimodal Large Language Models (MLLMs) have achieved remarkable success in tasks requiring complex cross-modal understanding and reasoning, leading to their rapid adoption across numerous applications~\cite{qwen3.5, gemini3.1pro, gpt5.2, hao2026unix}. 
Despite these advances, deploying MLLMs in knowledge-intensive scenarios still raises important reliability concerns. 
In particular, MLLMs may produce multimodal hallucinations and often lack access to private or domain-specific knowledge, since their parameters are primarily trained on large-scale public corpora~\cite{zhou2025they, wasserman2025real, huang2025survey}. 
To address these limitations, Multimodal Retrieval-Augmented Generation (MM-RAG) augments MLLMs with external knowledge bases by retrieving relevant multimodal evidence to ground the generation process~\cite{hu2023reveal, caffagni2024wiki, zhang2025comprehensive, zhao2026partially, you2026knowledge}. 
By incorporating factual context at inference time, MM-RAG significantly improves both the reliability and domain adaptability of MLLMs.

\begin{wrapfigure}{r}{0.55\columnwidth}
  \vspace{-1.75em}
  \centering
  \includegraphics[width=0.99\linewidth]{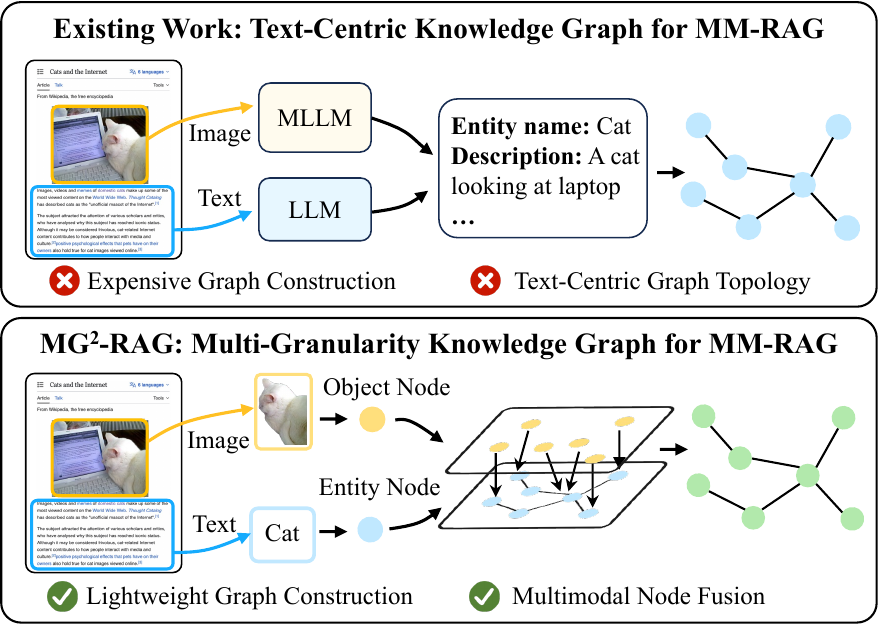}
  \vspace{-1.5em}
  \caption{\textbf{Comparison between existing graph-based MM-RAG frameworks and MG$^2$-RAG.} Existing methods rely on costly text-centric graph construction that discards fine-grained visual information, whereas MG$^2$-RAG efficiently fuses textual entities and visual objects into unified multimodal nodes while preserving atomic evidence.}
  \label{fig:head}
  \vspace{-1.75em}
\end{wrapfigure}

Existing MM-RAG methods generally fall into two paradigms: vector-based and graph-based approaches.
Vector-based MM-RAG encodes multimodal inputs into a shared embedding space~\cite{radford2021learning, sun2024eva, nussbaum2025nomic} and retrieves relevant evidence via similarity search~\cite{cocchi2025augmenting, compagnoni2026reag}. 
While effective in many scenarios, this paradigm typically retrieves isolated multimodal elements and largely ignores the structural relationships among different pieces of evidence~\cite{chen2022murag, yan2024echosight, lim2024unirag,  hu2025mrag}. 
Consequently, two fundamental limitations arise:
First, the semantic gap between modalities can exacerbate modality fragmentation, weakening cross-modal alignment. 
Second, similarity-based retrieval cannot model logical dependencies among retrieved evidence, making complex multi-hop reasoning difficult.

To overcome these limitations, recent work has explored graph-based MM-RAG, which organizes multimodal knowledge into structured graphs with explicit relational dependencies. 
By representing entities and their relationships as graph nodes and edges, these approaches enable retrieval through structured traversal rather than flat similarity matching~\cite{yuan2025mkg, wan2025mmgraphrag, liu2025aligning, you2026cut}. 
Such structured representations allow models to reason over interconnected multimodal facts and improve reliability in knowledge-intensive tasks~\cite{zhuang2026linearrag, gutierrez2025from, guo2024lightrag}. 
As a result, graph-based retrieval has emerged as a promising direction for bridging multimodal alignment and structured reasoning.
Despite its promise, existing graph-based MM-RAG systems still face several practical challenges.
\begin{itemize}[nolistsep]
  \item \textbf{Expensive Graph Construction:} 
  Many existing methods rely heavily on MLLMs to extract relational triplets from multimodal data, resulting in significant overhead when processing large-scale knowledge bases \cite{liu2025aligning, yuan2025mkg, wan2025mmgraphrag}. 

  \item \textbf{Text-Centric Graph Topology:}
  Visual information is often converted into textual descriptions before constructing the graph~\cite{guo2024lightrag, zhuang2026linearrag, wan2025mmgraphrag}. 
  This transformation discards fine-grained visual structures and overlooks independent visual entities that may contain important semantic information.

  \item \textbf{Difficult Cross-Modal Retrieval:}
  Since these graphs lack unified multimodal concepts, they struggle to effectively process queries containing both text and images, which limits semantic propagation across modalities and weakens multi-hop reasoning capabilities~\cite{liu2025aligning, yan2024echosight}.
\end{itemize}

To address these challenges, we propose \textbf{MG$^2$-RAG}, a lightweight \textbf{M}ulti-\textbf{G}ranularity \textbf{G}raph \textbf{RAG} framework for efficient and reliable multimodal retrieval-augmented generation.
MG$^2$-RAG constructs hierarchical multimodal knowledge graphs by combining lightweight textual parsing with entity-driven visual grounding, avoiding expensive MLLM-driven graph construction.
It further fuses textual entities and visual objects into unified multimodal nodes while preserving fine-grained atomic evidence from both modalities.

Building upon this representation, we introduce a multi-granularity graph retrieval mechanism that aggregates dense similarities onto multimodal nodes and propagates relevance through the graph topology, enabling unified multimodal retrieval and structured multi-hop reasoning.
Extensive experiments on four representative multimodal tasks, including retrieval, knowledge-based VQA, reasoning, and classification, demonstrate that MG$^2$-RAG consistently achieves state-of-the-art performance while substantially reducing graph construction overhead.
Our main contributions are summarized as follows:
\begin{itemize}[nolistsep]
  \item \textbf{Lightweight Multimodal Knowledge Graph Construction.}
  We propose a lightweight pipeline that bypasses the expensive MLLM-driven triplet extraction process, significantly reducing the time and cost of multimodal knowledge graph construction.

  \item \textbf{Modality-Preserving Multimodal Node Fusion.}
  We introduce a modality-preserving fusion strategy that integrates textual entities and visual objects into unified multimodal nodes while preserving fine-grained atomic evidence from both modalities.

  \item \textbf{Multi-Granularity Graph Retrieval.}
  We design a retrieval mechanism that aggregates dense similarities onto multimodal nodes and propagates relevance through the graph topology, enabling unified multimodal retrieval and multi-hop reasoning for MLLM generation.
\end{itemize}

\section{Related Work}
\label{sec:related}

\vspace{-0.5em}
\paragraph{RAG for LLMs}
RAG has evolved from static \emph{retrieve-then-generate} pipelines to \emph{dynamic, agentic reasoning} systems. 
Early approaches mitigate hallucinations by augmenting language models with external knowledge, progressing from sparse Wikipedia retrieval~\cite{chen2017reading, lee2019latent} to dense retrieval~\cite{guu2020retrieval, lewis2020retrieval, lazaridou2022internet}.
Subsequent work introduces adaptive retrieval strategies that enable models to determine when and what to retrieve, often optimized through reinforcement learning and environmental feedback \cite{asai2024selfrag, cheng2024unified, gao2025smartrag, hsu2025grounding}.
Recently, RAG has become tightly coupled with multi-hop reasoning through hierarchical query decomposition~\cite{lee2024planrag, zhang2025levelrag}, iterative retrieval-reasoning loops~\cite{wang2025chain, guan2026deeprag}, and agentic workflows~\cite{li2025search, wu2025agentic, lee2025rearag}.
These advances improve performance on complex, knowledge-intensive tasks.
Nonetheless, most existing methods still rely on \emph{flat vector retrieval}, limiting their ability to model structured semantic dependencies and motivating graph-based retrieval paradigms.

\paragraph{Graph RAG}
To overcome the structural limitations of flat retrieval, recent work has incorporated graph-based representations and decoupled subgraph retrieval into RAG~\cite{zhang2022subgraph, edge2024local, sarthi2024raptor, he2024g, luo2025gfm, you2026knowledge}. 
Graph RAG frameworks leverage entity-level knowledge graphs and community structures to better capture global dependencies and multi-hop relationships~\cite{edge2024local, edge2024lazygraphrag, you2026cut}.
Given the high computational cost of graph construction, subsequent research has focused on scalable indexing and retrieval strategies, including dual-level retrieval~\cite{guo2024lightrag}, bipartite or relation-free graph designs~\cite{huang2025ket, zhuang2026linearrag}, lightweight parallel triple-scoring~\cite{li2025simple}, and hierarchical tree-based structures~\cite{zhao2025e2graphrag, wang2026archrag}. 
Beyond efficiency, integrating graphs with LLMs has proven crucial for faithful multi-hop reasoning and long-term memory modeling.
Iterative graph exploration ~\cite{sun2024think, jin2024graph, ma2025think, yu2025can}, planning over structured representations~\cite{chen2024plan, luo2024reasoning}, memory-inspired architectures~\cite{jimenez2024hipporag, gutierrez2025from}, and agentic frameworks~\cite{dong2026youtugraphrag} demonstrate that explicit relational modeling significantly enhances reasoning depth and interpretability. 
Nevertheless, these methods are predominantly \emph{text-centric} and are not directly designed for multimodal settings.

\paragraph{Multimodal RAG}
Vision-language pre-training models~\cite{radford2021learning, li2022blip} have significantly advanced multimodal understanding, yet they remain constrained by parametric knowledge. 
Multimodal RAG (MM-RAG) addresses this limitation by introducing external memory for grounding generation in retrieved visual–textual evidence~\cite{chen2022murag, yasunaga2022retrieval, hu2023reveal}.
Recent efforts improve retrieval granularity~\cite{faysse2025colpali, luo2025videorag} and incorporate adaptive or self-reflective strategies~\cite{cocchi2025augmenting, zhang2024mr2ag, caffagni2024wiki} to filter noise and control retrieval necessity. 
However, most systems rely on flat vector similarity, which struggles with structured multi-hop reasoning across modalities~\cite{krishna2017visual, liu2019mmkg}.

To address this, Multimodal Knowledge Graphs (MMKGs)~\cite{liu2025aligning, yuan2025mkg, lee2024multimodal} have emerged as a promising direction for structured multimodal retrieval.
For example, MMGraphRAG~\cite{wan2025mmgraphrag} leverages scene graphs and clustering to connect visual and textual information.
Yet existing methods typically rely on MLLMs to translate visual content into textual graph nodes, yielding text-centric graph representations that discard fine-grained visual structures.
Consequently, preserving atomic visual regions together with textual entities in a \textbf{natively unified multimodal graph} remains largely underexplored.

\section{Methodology}
\label{sec:method}

\begin{figure*}[t]
  \centering
  \includegraphics[width=0.99\textwidth]{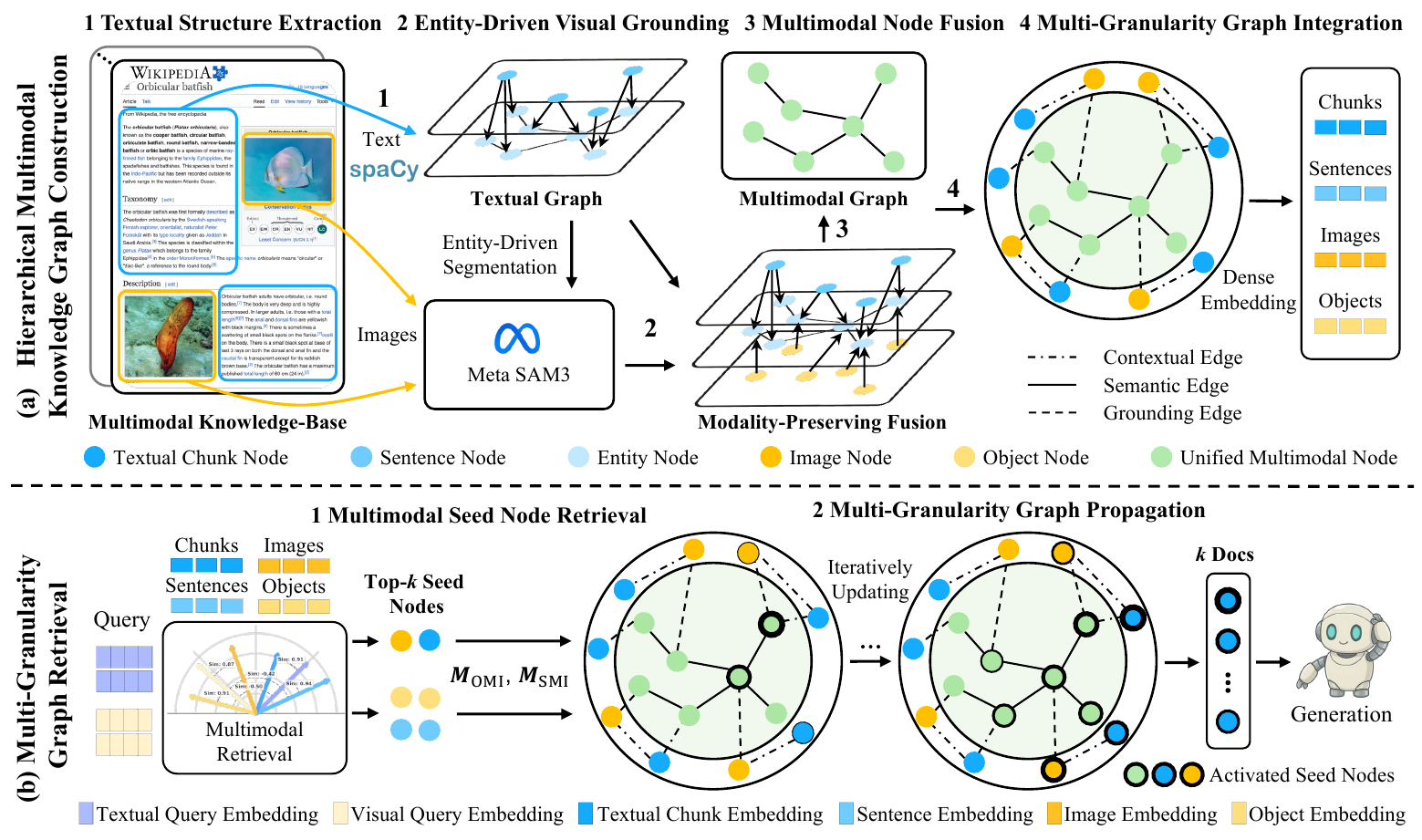}
  \vspace{-0.5em}
  \caption{\textbf{Overview of MG$^2$-RAG.} MG$^2$-RAG contains two modules: \textbf{(a) Hierarchical Multimodal Knowledge Graph Construction} that transforms a multimodal knowledge base into a hierarchical fine-grained multimodal knowledge graph (MMKG) via textual parsing and entity-driven visual grounding; \textbf{(b) Multi-Granularity Graph Retrieval} that employs a multi-granularity retrieval mechanism, driven by graph propagation, to retrieve the top-K most relevant document chunks for MLLM generation.}
  \label{fig:framework}
  \vspace{-1.0em}
\end{figure*}

We propose \textbf{MG$^2$-RAG}, a lightweight framework that transforms an unstructured multimodal knowledge base into a multi-granularity MMKG for retrieval-augmented generation.
As shown in Figure~\ref{fig:framework}, the framework consists of two modules:
(1) Hierarchical Multimodal Knowledge Graph Construction and
(2) Multi-Granularity Graph Retrieval.

\subsection{Hierarchical Multimodal Knowledge Graph Construction}
\label{sec:method:graph_construction}

\paragraph{Motivation}
Existing MMKG construction strategies suffer from two key issues:
(1) expensive LLM-based relation extraction and
(2) modality fragmentation caused by converting images into textual descriptions.
To address these issues, we construct a hierarchical MMKG that directly aligns textual entities with grounded visual objects, forming atomic unified multimodal nodes.
We define the multimodal knowledge base as $\mathcal{D} = \{(c_k, \mathcal{I}_k)\}_{k=1}^N$, where each unit consists of a textual document chunk $c_k$ and its associated image set $\mathcal{I}_k = \{I_{k,1}, \cdots, I_{k,m_k}\}$. 
Rather than modeling text–image correspondence as loose cross-modal links, we explicitly extract and fuse semantic information to construct a compact hierarchical MMKG.
Each unified multimodal node preserves fine-grained textual and visual components internally, enabling nuanced downstream reasoning.

\paragraph{Textual Structure Extraction: NER and Dependency Parsing}
We first extract structural signals from text using the transformer-based spaCy model \texttt{en\_core\_web\_trf} \cite{honnibal2020spacy}. 
Each chunk $c_k$ is parsed into a sentence set $\mathcal{S}_k = \{s_{k,1}, \cdots, s_{k,n_k}\}$, from which we extract a local entity set $\mathcal{E}_k$. 
Aggregating across all chunks yields the global sentence and entity sets:
\[
  \textstyle \mathcal{S} = \bigcup_{k=1}^N \mathcal{S}_k, \quad \mathcal{E} = \bigcup_{k=1}^N \mathcal{E}_k.
\]
To capture structural relations between entities, we leverage token-level dependency parsing.
Instead of relying on expensive LLM-based OpenIE extraction~\cite{jimenez2024hipporag, dong2026youtugraphrag}, we adopt a \emph{lightweight} rule-based relation extraction strategy grounded in dominant grammatical patterns (e.g., predicate-centered and nominal-modifier structures).
This significantly reduces construction cost while preserving high-precision relational structure. 
Details are provided in Appendix~\ref{app:impl:relation-extraction}.

\paragraph{Entity-Driven Visual Grounding}
To prevent the ``translation-to-text'' bottleneck, we directly ground textual entities in visual space.
Given the local entity set $\mathcal{E}_k$ and image set $\mathcal{I}_k$ for each chunk $c_k$, we perform \emph{entity-driven segmentation} rather than global image parsing. 
Specifically, we batch entity names as semantic prompts into an open-vocabulary segmentation model $\Phi$ (SAM3~\cite{carion2025sam}) to localize entity-relevant regions. 
For entity $e \in \mathcal{E}_k$, its grounded region set is:
\begin{equation}
  \mathcal{V}_{o}^e = \{ o \mid (o, \sigma) \in \Phi(e, I), \, I \in \mathcal{I}_k, \, \sigma > \tau \},
\end{equation}
where $o$ is the segmented object, $\sigma$ denotes grounding confidence for entity $e$ on image $I$, and $\tau$ is a threshold. 
This targeted grounding ensures that visual objects are directly and semantically aligned with textual entities.

\paragraph{Modality-Preserving Multimodal Node Fusion}
For entities with valid detections ($\mathcal{V}_{o}^e \neq \emptyset$), we construct a modality-preserving multimodal node $v_m = (e, \mathcal{V}_{o}^e) \in \mathcal{V}_M$, where $\mathcal{V}_M$ denotes the set of multimodal nodes.
To explicitly model internal composition, we define:
\begin{itemize}[nolistsep]
  \item \textbf{Object-Multimodal Incidence Matrix:} 
  $\bm{M}_{\text{OMI}} \in \{0,1\}^{|\mathcal{V}_O| \times |\mathcal{V}_M|}$, where $\mathcal{V}_O = \bigcup_{e \in \mathcal{E}} \mathcal{V}_{o}^e$. 
  Entry $\bm{M}_{\text{OMI}}[i,j]=1$ indicates that object $o_i \in \mathcal{V}_O$ belongs to multimodal node $v_{m_j} \in \mathcal{V}_M$.

  \item \textbf{Sentence-Multimodal Incidence Matrix:} 
  $\bm{M}_{\text{SMI}} \in \{0,1\}^{|\mathcal{S}| \times |\mathcal{V}_M|}$, 
  where $\bm{M}_{\text{SMI}}[i,j] = 1$ indicates entity $e_j$ that is associated with the unified multimodal node $v_{m_j}$ appears in sentence $s_i \in \mathcal{S}$.
\end{itemize}
These matrices enable structured aggregation during retrieval.

\paragraph{Multi-Granularity Graph Integration}
Building upon the matrices of document chucks $\mathcal{V}_C$, images $\mathcal{V}_I$, and multimodal nodes $\mathcal{V}_M$, we construct multi-granularity multimodal graph:
\[
  \mathcal{G} = (\mathcal{V}, \mathcal{E}_\mathcal{G}), \quad \mathcal{V} = \mathcal{V}_C \cup \mathcal{V}_I \cup \mathcal{V}_M.
\]
For each multimodal unit $(c_k, \mathcal{I}_k)$, nodes operate at three levels: chunk nodes $v_c^k \in \mathcal{V}_C$, image nodes $v_I^{k,j} \in \mathcal{V}_I$ for each associated image $I_{k,j} \in \mathcal{I}_k$, and unified multimodal nodes $v_m \in \mathcal{V}_M$.
To represent rich structural and semantic interactions, the edge set $\mathcal{E}_\mathcal{G}$ is constructed across three complementary dimensions: 
\begin{enumerate}[nolistsep,label*=(\arabic*)]
  \item \textbf{Contextual Edges:} To preserve hierarchical provenance, each chunk node $v_c^k$ is linked to its associated image nodes $v_I^{k,j}$ and multimodal nodes $v_m$ derived from $e \in \mathcal{E}_k$.
  These edges maintain document-level structural coherence.
  
  \item \textbf{Semantic Edges:} To encode linguistic structure, we connect multimodal nodes $(v_{m_a}, v_{m_b})$ when their underlying textual entities exhibit direct syntactic dependencies. 
  This enables structured reasoning over explicit relational cues extracted from text.
  
  \item \textbf{Grounding Edges:} To model explicit cross-modal alignment, each visual object $o \in \mathcal{V}_o^e$ encapsulated within a multimodal node $v_m$ is linked to its source image node $v_I^{k,j}$. 
  The edge weight is assigned as the grounding confidence $\sigma$, reflecting the reliability of the localized visual evidence.
\end{enumerate}
These three edge types integrate contextual hierarchy, semantic structure, and visual grounding into a unified heterogeneous topology, laying the foundation for structured multi-granularity retrieval in Section \ref{sec:method:mm_retrieval}

In parallel, all textual and visual elements are embedded into a shared space via a unified encoder $\Psi(\cdot)$ (i.e., \texttt{EVA-CLIP-8B} \cite{sun2024eva}),\footnote{Without specialization, all embeddings are explicitly $\ell_2$-normalized, and their semantic proximity is computed directly via cosine similarity.} which maps any elemental input $x$ to a $d$-dimensional vector $\bm{z}_x = \Psi(x) \in \mathbb{R}^d$.
We construct four embedding matrices corresponding to distinct levels of granularity: 
(1) \textbf{Sentence matrix} $\bm{Z}_S \in \mathbb{R}^{|\mathcal{S}| \times d}$ encoded from global sentences $\mathcal{S}$; 
(2) \textbf{Chunk matrix} $\bm{Z}_C \in \mathbb{R}^{|\mathcal{V}_C| \times d}$ encoded from document chunks $\mathcal{V}_C$; 
(3) \textbf{Image matrix} $\bm{Z}_I \in \mathbb{R}^{|\mathcal{V}_I| \times d}$ encoded from source images $\mathcal{V}_I$; and 
(4) \textbf{Object matrix} $\bm{Z}_O \in \mathbb{R}^{|\mathcal{V}_O| \times d}$ encoded from segmented objects $\mathcal{V}_O$.
Together, the graph topology and dense embeddings enable multi-granularity retrieval.
\vspace{-0.25em}

\subsection{Multi-Granularity Graph Retrieval}
\label{sec:method:mm_retrieval}

\vspace{-0.25em}
\paragraph{Motivation}
Flat vector retrieval treats modalities independently and ignores structural dependencies within multimodal knowledge bases. 
To enable structured cross-modal reasoning, we adopt a two-stage strategy: 
(1) activating fine-grained multimodal evidence as seed nodes, and 
(2) propagating their relevance across the graph topology to capture multi-hop interactions.

\paragraph{Multimodal Seed Node Retrieval}
Leveraging the multimodal index constructed in Section~\ref{sec:method:graph_construction}, a textual query $q_t$ or visual query $q_v$ retrieves candidate evidence from all semantic levels. 
For each modality $m \in \{t,v\}$, dense similarities with the embedding matrices $\bm{Z}_S$, $\bm{Z}_C$, $\bm{Z}_I$, and $\bm{Z}_O$ produce four relevance vectors:
$\bm{s}_S^{(m)} \in \mathbb{R}^{|\mathcal{S}|}$,
$\bm{s}_C^{(m)} \in \mathbb{R}^{|\mathcal{V}_C|}$,
$\bm{s}_I^{(m)} \in \mathbb{R}^{|\mathcal{V}_I|}$, and
$\bm{s}_O^{(m)} \in \mathbb{R}^{|\mathcal{V}_O|}$.

Because sentences $\mathcal{S}$ and visual objects $\mathcal{V}_O$ act as intermediate pivots rather than primary graph nodes, their scores are aggregated onto unified multimodal nodes $\mathcal{V}_M$ using the incidence matrices $\bm{M}_{\text{SMI}}$ and $\bm{M}_{\text{OMI}}$:
\begin{equation}
\bm{s}_{M}^{(m)} =
\bm{D}_S^{-1}\bm{M}_{\text{SMI}}^\top \bm{s}_S^{(m)}
+
\bm{D}_O^{-1}\bm{M}_{\text{OMI}}^\top \bm{s}_O^{(m)},
\end{equation}
where $\bm{D}_S, \bm{D}_O \in \mathbb{R}^{|\mathcal{V}_M| \times |\mathcal{V}_M|}$ are diagonal degree matrices performing mean pooling based on the number of mapped unified multimodal nodes. 

To balance evidence from different semantic levels, we introduce scaling factors $\omega_C$ and $\omega_I$ for chunk and image nodes. 
Given the node partition $\mathcal{V} = \mathcal{V}_C \cup \mathcal{V}_I \cup \mathcal{V}_M$, the modality-specific activation vector is $\bm{u}^{(m)} = [\,\omega_C \bm{s}_C^{(m)};\ \omega_I \bm{s}_I^{(m)};\ \bm{s}_M^{(m)}\,]$.
Finally, modality contributions are fused using weights $\lambda_t$ and $\lambda_v$:
\begin{equation}
  \bm{r}_{0} = \lambda_t \bm{u}^{(t)} + \lambda_v \bm{u}^{(v)}.
\end{equation}
To reduce noise before graph propagation, we retain only the top-$k$ activated nodes and normalize the resulting distribution, as detailed in Appendix~\ref{app:iml:task-setting}.

\paragraph{Multi-Granularity Graph Propagation}
Starting from the seed distribution $\bm{r}_{0}$, we propagate relevance across the heterogeneous graph using Personalized PageRank~\cite{haveliwala2002topic}. 
Let $\bm{W}$ denote the column-normalized transition matrix derived from $\mathcal{E}_\mathcal{G}$.
The diffusion process iteratively updates node activations as:
\begin{equation}
  \bm{r}_{\ell+1} = \alpha \bm{W}\bm{r}_{\ell} + (1-\alpha)\bm{r}_{0},
\end{equation}
until $\Delta \bm{r} = \bm{r}_{\ell+1} - \bm{r}_{\ell} \leq \epsilon$, where $\alpha \in (0, 1)$ controls the propagation strength and $(1-\alpha)$ ensures periodic restarts from the seed distribution $\bm{r}_{0}$.
After convergence, the relevance scores of chunk nodes are extracted to select the top-$k$ document chunks. 
These chunks are combined with the multimodal query $(q_t, q_v)$ to form the final prompt for the MLLM, enabling grounded response generation.

\section{Experiment}
\label{sec:expt}

\subsection{Experimental Setup}
\label{sec:expt:setup}

\vspace{-0.5em}
\paragraph{Baselines}
We compare \textbf{MG$^2$-RAG} against three state-of-the-art graph-based MM-RAG frameworks: \textbf{VaLiK}~\cite{liu2025aligning}, \textbf{mKG-RAG}~\cite{yuan2025mkg}, and \textbf{MMGraphRAG}~\cite{wan2025mmgraphrag}. 
For fair comparison, we report results using the best-performing configurations from the original papers without dataset-specific fine-tuning. 
Additional comparisons with other relevant approaches are summarized in Tables \ref{tab:retrieval}--\ref{tab:crisismmd}.

\paragraph{Multimodal Tasks, Datasets, and Evaluation Metrics} 
We evaluate MG$^2$-RAG on four representative multimodal tasks, covering retrieval, knowledge-based Visual Question Answering (VQA), reasoning, and classification.
\begin{itemize}[nolistsep]
  \item \textbf{Multimodal Retrieval:} 
  We evaluate retrieval performance on two Wikipedia-linked datasets: \textbf{E-VQA} \cite{mensink2023encyclopedic} and \textbf{InfoSeek} \cite{chen2023can}. 
  E-VQA contains diverse fine-grained entities and includes \textbf{single-hop} and \textbf{two-hop} multi-page queries, making retrieval challenging. 
  InfoSeek provides two evaluation splits: Unseen Entity (\textbf{Unseen-E}) and Unseen Question (\textbf{Unseen-Q}), where the former requires retrieving previously unseen entities.
  For both datasets, we construct MMKGs from the associated corpora and retrieve supporting evidence for each query. 
  Performance is evaluated using Recall@K (\textbf{R@K}).

  \item \textbf{Knowledge-based VQA:} 
  Using the retrieved evidence, the model generates answers for both textual and visual queries. 
  In \textbf{E-VQA}, questions may involve multiple associated images and require reasoning across one or two knowledge hops. 
  For \textbf{InfoSeek}, evaluation on both Unseen-E and Unseen-Q assesses generalization to novel entities and question formulations. 
  Performance is measured using BERT Matching Score (\textbf{BEM}) \cite{bulian-etal-2022-tomayto} for E-VQA, and \textbf{standard accuracy}~\cite{goyal2017making} or \textbf{relaxed accuracy}~\cite{methani2020plotqa} for InfoSeek.
  
  \item \textbf{Multimodal Reasoning:} 
  We further evaluate our method on \textbf{ScienceQA} \cite{lu2022learn}, which contains around 21k multimodal science questions spanning physics, chemistry, and biology.
  Following VaLiK \cite{liu2025aligning}, the training set is used as the knowledge base to construct the MMKG. 
  The task requires answering complex scientific questions through multimodal reasoning. Performance is measured using \textbf{classification accuracy}, reported across different question types, contextual modalities, and educational stages.

  \item \textbf{Multimodal Classification:} 
  To assess real-world applicability, we evaluate MG$^2$-RAG on \textbf{CrisisMMD} \cite{alam2018crisismmd}, a disaster-response dataset containing approximately 35k noisy social-media image-text pairs. 
  Following VaLiK \cite{liu2025aligning}, the training set is used to construct the MMKG, capturing modality correlations and noise patterns typical of user-generated content.
  We evaluate three classification tasks: 
  (1) \textbf{BC}: binary classification for information relevance filtering, 
  (2) \textbf{MC}: multi-class classification of fine-grained humanitarian categories, and 
  (3) \textbf{MC-m}: a simplified multi-class setting with merged categories.
  Performance is measured using \textbf{standard classification accuracy}.
\end{itemize}
Further details on dataset preprocessing are provided in Appendix \ref{app:impl:dataset}.

\paragraph{Implementation Details}
For \textbf{Hierarchical MMKG Construction}, textual analysis is performed using spaCy (\texttt{en\_core\_web\_trf}) \cite{honnibal2020spacy}, and entity-driven visual grounding is conducted with SAM3~\cite{carion2025sam}.
All textual and visual elements are embedded into a shared semantic space using \texttt{EVA-CLIP-8B}~\cite{sun2024eva}. 
For \textbf{Multi-Granularity Graph Retrieval}, both dense similarity computation and Personalized PageRank (PPR) propagation are implemented on GPUs to accelerate large-scale inference. 
Hyperparameters are empirically tuned for each task, with detailed configurations and prompts provided in Appendix \ref{app:iml:task-setting}.
All experiments are conducted on two NVIDIA RTX 6000 Ada GPUs (48GB).

\subsection{Graph Construction Efficiency}
\label{sec:expt:construction}

\begin{figure*}[t]
  \centering
  \includegraphics[width=0.99\textwidth]{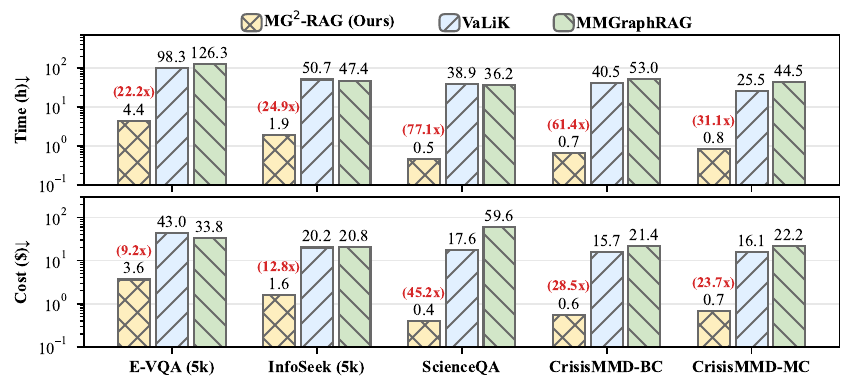}
  \vspace{-1.0em}
  \caption{\textbf{Graph construction efficiency.} The top and bottom plots show construction time (hours) and cost (\$), respectively. Red annotations above the MG$^2$-RAG bars indicate the efficiency gain (speedup or cost reduction) relative to the strongest baseline.}
  \label{fig:efficiency}
  \vspace{-1.5em}
\end{figure*}

\paragraph{Construction Efficiency}
To evaluate the efficiency of our hierarchical MMKG construction strategy, we compare the graph construction time and cost of MG$^2$-RAG with existing open-source graph-based MM-RAG baselines. 
For large benchmarks (E-VQA and InfoSeek), we restrict the knowledge base to a 5k subset because baseline approaches rely on expensive MLLM-driven triplet extraction, making full-scale graph construction computationally and financially infeasible. 
In contrast, MG$^2$-RAG remains efficient enough to construct graphs over the entire knowledge base, highlighting its scalability advantage.

Figure \ref{fig:efficiency} summarizes the results. 
By bypassing the bottleneck of MLLM-based relation extraction and instead combining lightweight textual parsing with entity-driven visual grounding, MG$^2$-RAG achieves substantial efficiency gains across all evaluated datasets. The calculation method is detailed in Appendix \ref{app:impl:dataset}.
\begin{itemize}[nolistsep]
  \item \textbf{Construction Time:} MG$^2$-RAG dramatically accelerates graph construction. Compared to the strongest baseline for each dataset, our framework achieves a maximum speedup of \textbf{77.1$\times$} on ScienceQA and an average speedup of \textbf{43.3$\times$} across all benchmarks (ranging from 22.2$\times$ to 77.1$\times$).
  
  \item \textbf{Construction Cost:} The lightweight strategy also significantly reduces financial overhead. MG$^2$-RAG lowers construction costs by up to \textbf{45.2$\times$} on ScienceQA, with an average cost reduction of \textbf{23.9$\times$} across the five datasets.
\end{itemize}

These results demonstrate that our MMKG construction strategy effectively tackles a major scalability bottleneck of existing graph-based MM-RAG methods, enabling practical deployment without sacrificing downstream performance.

\paragraph{Scalability Results}
We further evaluate the scalability of MG$^2$-RAG on E-VQA by increasing the knowledge base from 1k to 100k documents.
As shown in Table~\ref{tab:camera_scalability}, graph construction scales nearly linearly, with processing time increasing from \textbf{0.6} to \textbf{63.4 hours}. Under the largest setting, using two GPUs further reduces the construction time to \textbf{32.8 hours}, demonstrating that MG$^2$-RAG scales efficiently to large multimodal knowledge bases.

\begin{table}[t]
\centering
\small
\setlength{\tabcolsep}{6pt}
\renewcommand{\arraystretch}{1.0}
\caption{\textbf{Scalability of graph construction on E-VQA.} We report the construction time under different knowledge-base sizes. $100\mathrm{k}^{*}$ denotes the setting with double GPUs.}
\vspace{-0.5em}
\label{tab:camera_scalability}
\resizebox{0.6\columnwidth}{!}{%
\begin{tabular}{lcccccc}
  \toprule
  \textbf{Data Size ($N$)} & 1k & 5k & 10k & 50k & 100k & $100\mathrm{k}^{*}$ \\
  \midrule
  \textbf{Time (Hours)} & 0.6 & 2.9 & 5.7 & 29.7 & 63.4 & 32.8 \\
  \bottomrule
\end{tabular}}
\end{table}
\setlength{\textfloatsep}{1.0em}

\subsection{Multimodal Retrieval Performance}
\label{sec:expt:retrieval}

\begin{table*}[!t]
\centering
\small 
\setlength{\tabcolsep}{5pt} 
\renewcommand{\arraystretch}{1.2} 
\caption{\textbf{Retrieval performance on E-VQA and InfoSeek.} Retrieval modes include \textbf{T$\rightarrow$V} (Text-to-Vision), \textbf{V$\rightarrow$T} (Vision-to-Text), and \textbf{MM} (Multimodal retrieval). \textbf{Bold} and \underline{underlined} denote the best and second-best results.}
\vspace{-0.5em}
\label{tab:retrieval}
\resizebox{0.99\textwidth}{!}{
\begin{tabular}{ll rrrrr rrrrr}
    \toprule
    \multirow{2}{*}[-0.75ex]{\textbf{Model}} & \multirow{2}{*}[-0.75ex]{\textbf{\shortstack{Retrieval \\ Modality}}} & \multicolumn{5}{c}{\textbf{E-VQA}} & \multicolumn{5}{c}{\textbf{InfoSeek}} \\
    \cmidrule(lr){3-7} \cmidrule(lr){8-12}
     & & \textbf{R@1$\uparrow$} & \textbf{R@5$\uparrow$} & \textbf{R@10$\uparrow$} & \textbf{R@20$\uparrow$} & \textbf{R@50$\uparrow$} 
     & \textbf{R@1$\uparrow$} & \textbf{R@5$\uparrow$} & \textbf{R@10$\uparrow$} & \textbf{R@20$\uparrow$} & \textbf{R@50$\uparrow$} \\
    \midrule
    \rowcolor[HTML]{D9EAD3}
    \multicolumn{12}{c}{\textbf{\textit{Single Modality Retrieval}}} \\
    \multirow{2}{*}{\textbf{Nomic-Embed \cite{nussbaum2025nomic}}} 
     & T$\rightarrow$T & 4.7 & 9.1 & 12.1 & 15.9 & 23.0 & 0.2 & 1.0 & 1.5 & 3.0 & 5.6 \\
     & V$\rightarrow$V & 16.4 & 26.9 & 33.6 & 38.4 & 44.5 & 31.4 & 48.8 & 54.5 & 60.0 & 66.1 \\
    \midrule
    \multirow{2}{*}{\textbf{CLIP ViT-L/14 \cite{radford2021learning}}} 
     & T$\rightarrow$T & 1.0 & 2.0 & 2.5 & 3.2 & 4.4 & 0.1 & 0.2 & 0.2 & 0.4 & 0.7 \\
     & V$\rightarrow$V & 17.8 & 31.0 & 36.4 & 42.6 & 50.3 & 32.7 & 52.3 & 59.6 & 66.2 & 74.0 \\
    \midrule
    \multirow{2}{*}{\textbf{EVA-CLIP-8B \cite{sun2024eva}}} 
     & T$\rightarrow$T & 2.2 & 4.1 & 5.2 & 7.0 & 9.5 & 0.1 & 0.7 & 1.4 & 2.3 & 4.5 \\
     & V$\rightarrow$V & 30.0 & 45.0 & 50.0 & 55.0 & 59.9 & 47.2 & 65 & 70.2 & 74.4 & 79.6 \\
     \midrule
    \rowcolor[HTML]{E7F3FF}
    \multicolumn{12}{c}{\textbf{\textit{Cross-Modality Retrieval}}} \\
    \multirow{2}{*}{\textbf{CLIP ViT-L/14 \cite{radford2021learning}}} 
     & T$\rightarrow$V & 1.6 & 3.3 & 4.8 & 7.0 & 10.9 & 0.2 & 0.6 & 1.1 & 2.3 & 5.3 \\
     & V$\rightarrow$T & 10.9 & 26.2 & 35.8 & 44.6 & 55.6 & 20.8 & 40.7 & 49.2 & 57.6 & 67.7 \\
    \midrule
    
    \multirow{2}{*}{\textbf{EVA-CLIP-8B \cite{sun2024eva}}} 
     & T$\rightarrow$V & 1.8 & 4.6 & 6.3 & 8.5 & 12.9 & 0.2 & 0.4 & 0.8 & 2.0 & 4.9 \\
     & V$\rightarrow$T & \underline{42.0} & \underline{62.6} & \underline{69.5} & \underline{75.6} & \underline{83.2} & \underline{56.5} & \underline{76.3} & \underline{82.2} & \underline{86.4} & \underline{90.6} \\
    \midrule

    \rowcolor[HTML]{FFF2CC}
    \multicolumn{12}{c}{\textbf{\textit{Multi-Modality Retrieval}}} \\
    \textbf{mKG-RAG \cite{yuan2025mkg}} & \centering{MM} & -- & -- & -- & -- & -- & 49.7 & 71.6 & 78.0 & 82.5 & 89.1 \\
    \textbf{MG$^2$-RAG (Ours)} & \centering{MM} & \textbf{44.9} & \textbf{66.4} & \textbf{72.0} & \textbf{79.4} & \textbf{83.9} & \textbf{59.6} & \textbf{78.9} & \textbf{83.8} & \textbf{87.4} & \textbf{91.0} \\    
    \bottomrule
\end{tabular}}
\end{table*}

Table \ref{tab:retrieval} reports retrieval results on E-VQA and InfoSeek, comparing MG$^2$-RAG with both cross-modal retrieval models and graph-based baselines.
Our framework consistently achieves the highest recall across all metrics (R@1--R@50).  
For instance, MG$^2$-RAG reaches R@1 scores of \textbf{44.9} and \textbf{59.6} on E-VQA and InfoSeek, respectively, as well as R@10 scores of \textbf{72.0} and \textbf{83.8}. 
These results surpass the strongest cross-modal baseline EVA-CLIP-8B ($V \to T$), which obtains 42.0 / 56.5 at R@1 and 69.5 / 82.2 at R@10. 

Compared with the multimodal graph baseline mKG-RAG, our framework achieves a further \textbf{+9.9} improvement at R@1 on InfoSeek.
These gains directly validate the effectiveness of our \textbf{multi-granularity graph retrieval mechanism}. 
By aggregating dense similarities onto unified multimodal nodes, MG$^2$-RAG bridges textual and visual evidence within a shared graph structure, allowing retrieval to capture cross-modal dependencies and structural relationships, thereby retrieving more accurate evidence than modality-isolated approaches.

\subsection{Knowledge-based VQA Performance}
\label{sec:expt:KBVQA}

\begin{table*}[!t]
\centering
\small
\setlength{\tabcolsep}{5pt} 
\renewcommand{\arraystretch}{1.1} 
\caption{\textbf{Knowledge-based VQA performance on InfoSeek and E-VQA.} \textbf{T} and \textbf{V} denote Text and Vision modalities, respectively.}
\vspace{-0.5em}
\label{tab:vqa}
\resizebox{\textwidth}{!}{
\begin{tabular}{ll rrrrr}
  \toprule
  \multirow{2}{*}[-0.5ex]{\textbf{Model}} & \multirow{2}{*}[-0.5ex]{\textbf{LLM/MLLM}} & \multicolumn{2}{c}{\textbf{E-VQA}} & \multicolumn{3}{c}{\textbf{InfoSeek}} \\
  \cmidrule(lr){3-4} \cmidrule(lr){5-7}
  & & \textbf{Single-Hop$\uparrow$} & \textbf{All$\uparrow$} & \textbf{Unseen-Q$\uparrow$} & \textbf{Unseen-E$\uparrow$} & \textbf{All$\uparrow$} \\
  \midrule
  
  \rowcolor[HTML]{D9EAD3}
  \multicolumn{7}{c}{\textbf{\textit{Zero-Shot MLLMs}}} \\
  \textbf{InstructBLIP \cite{dai2023instructblip}} & Flan-T5XL & 11.90 & 12.00 & 8.90 & 7.40 & 8.10 \\
 
  \textbf{LLaVA-v1.5 \cite{liu2024improved}} & LLaMA-3.1-8B & 16.00 & 16.90 & 8.30 & 8.90 & 7.80 \\
 
  \textbf{Qwen2.5-VL-7B \cite{compagnoni2026reag}} & Qwen2.5-VL-7B & 23.60 & 23.20 & 22.80 & 24.10 & 23.70 \\
  \textbf{Gemini-3.1-Pro \cite{gemini3.1pro}} & Gemini-3.1-Pro & 29.36 & 27.90 & 29.92 & 22.52 & 25.70 \\
  \textbf{GPT-5.2 \cite{gpt5.2}} & GPT-5.2 & 44.19 & 40.30 & 32.68 & 26.54 & 29.29 \\
  \textbf{Qwen3.5-27B \cite{qwen3.5}} & Qwen3.5-27B & 29.79 & 25.72 & 20.25 & 16.75 & 18.33 \\
  \midrule
  
  \rowcolor[HTML]{E7F3FF} 
  \multicolumn{7}{c}{\textbf{\textit{Retrieval-Augmented Models}}} \\
  \textbf{RORA-VLM \cite{qi2024rora}} & Vicuna-7B & -- & -- & 25.10 & 27.30 & -- \\
  \textbf{EchoSight \cite{yan2024echosight} (V$\rightarrow$T)} & LLaMA-3.1-8B & 52.96 & 47.23 & 30.00 & 30.70 & 30.40 \\
  \textbf{EchoSight \cite{yan2024echosight} (V$\rightarrow$V)}& LLaMA-3.1-8B & 46.32 & 41.29 & 18.00 & 19.80 & 18.80 \\

  \midrule

  \rowcolor[HTML]{FFF2CC} 
  \multicolumn{7}{c}{\textbf{\textit{Graph Retrieval-Augmented Models}}} \\
  \textbf{mKG-RAG \cite{yuan2025mkg}} & LLaMA-3.1-8B & -- & -- & 32.90 & 31.30 & 32.10 \\

  \textbf{{MG}$^2$-RAG(Ours)} & LLaMA-3.1-8B & 55.57 & 47.77 & 32.32 & 32.84 & 32.58 \\
  \textbf{{MG}$^2$-RAG(Ours)} & Qwen2.5-VL-7B & 57.33 & 48.59 & 35.78 & 35.18 & 35.48 \\
  \textbf{{MG}$^2$-RAG(Ours)} & Qwen3.5-27B & 61.65 & 52.19 & 37.36 & 38.40 & 37.87 \\
  \textbf{{MG}$^2$-RAG(Ours)} & GPT-5.2 & \underline{68.96} & \textbf{60.30} & \underline{40.16} & \underline{38.47} & \underline{39.30} \\
  \midrule
  \textbf{VaLiK (5k) \cite{liu2025aligning}} & Qwen2.5-VL-7B & 16.16 & 15.22 & 3.19 & 2.07 & 2.51 \\
  \textbf{MMGraphRAG (5k) \cite{wan2025mmgraphrag}} & Qwen2.5-VL-7B & 19.12 & 16.52 & 0.69 & 0.39 & 0.50 \\
  \textbf{\textbf{{MG}$^2$-RAG (5k) (Ours)}} & Qwen2.5-VL-7B & 62.88 & 53.36 & 39.17 & 38.15 & 38.65 \\
  \textbf{\textbf{{MG}$^2$-RAG (5k) (Ours)}} & Qwen3.5-27B & \textbf{70.27} & \underline{60.24} & \textbf{42.49} & \textbf{42.03} & \textbf{42.26} \\
  \bottomrule
\end{tabular}}
\end{table*}

Table \ref{tab:vqa} presents results on knowledge-based VQA, where models must reason over retrieved multimodal evidence. 
MG$^2$-RAG establishes a new state-of-the-art, consistently outperforming zero-shot MLLMs, standard RAG methods, and leading graph-based approaches. 
Notably, with the Qwen3.5-27B backbone, MG$^2$-RAG attains \textbf{60.24} on E-VQA (5k) and \textbf{42.26} on InfoSeek (5k). Moreover, MG$^2$-RAG also benefits strong closed-source models. For example, with GPT-5.2, the E-VQA score improves from \textbf{40.30} to \textbf{60.30}, showing that retrieval still provides useful knowledge even when the base MLLM is strong.

Two key advantages explain these improvements.
First, compared with standard retrieval models like EchoSight \cite{yan2024echosight}, MG$^2$-RAG consistently yields better results, achieving +0.54 accuracy on E-VQA and +2.18 on InfoSeek.
This improvement highlights the effectiveness of our \textbf{modality-preserving multimodal node fusion}, which retains both textual entities and visual objects as unified concepts. 

Second, MG$^2$-RAG outperforms existing graph-based approaches even when using the same backbone models. 
For instance, with the LLaMA-3.1-8B backbone, our framework already surpasses mKG-RAG by 0.48 on InfoSeek, and the advantage becomes larger with stronger models. 
This demonstrates that our \textbf{lightweight MMKG construction strategy} produces a cleaner and more informative graph topology than traditional MLLM-generated triplet graphs. 

\subsection{Multimodal Reasoning Performance}
\label{sec:expt:multimodal-reasoning}

\begin{table*}[!t]
\centering
\small 
\setlength{\tabcolsep}{5pt} 
\renewcommand{\arraystretch}{1.1} 
\caption{\textbf{Multimodal reasoning performance on ScienceQA.} Results are reported across different domains and contexts: \textbf{NAT} (natural science), \textbf{SOC} (social science), \textbf{LAN} (language science), \textbf{TXT} (text context), \textbf{IMG} (image context), \textbf{NO} (no context), \textbf{G1-6} (grades 1-6), and \textbf{G7-12} (grades 7-12).}
\vspace{-0.5em}
\label{tab:scienceqa}
\resizebox{\textwidth}{!}{
\begin{tabular}{ll rrr rrr rrr}
  \toprule
  \multirow{2}{*}[-0.5ex]{\textbf{Model}} & \multirow{2}{*}[-0.5ex]{\textbf{LLM/MLLM}} & \multicolumn{3}{c}{\textbf{Subject}} & \multicolumn{3}{c}{\textbf{Context Modality}} & \multicolumn{2}{c}{\textbf{Grade}} & \multirow{2}{*}[-0.5ex]{\textbf{Avg.$\uparrow$}} \\
  \cmidrule(lr){3-5} \cmidrule(lr){6-8} \cmidrule(lr){9-10} 
  & & \textbf{NAT$\uparrow$} & \textbf{SOC$\uparrow$} & \textbf{LAN$\uparrow$} & \textbf{TXT$\uparrow$} & \textbf{IMG$\uparrow$} & \textbf{NO$\uparrow$} & \textbf{G1-6$\uparrow$} & \textbf{G7-12$\uparrow$} & \\
  \midrule
  \textbf{Human \cite{lu2022learn}} & - & 90.23 & 84.97 & 87.48 & 89.60 & 87.50 & 88.10 & 91.59 & 82.42 & 88.40 \\
  \midrule
  \rowcolor[HTML]{D9EAD3} \multicolumn{11}{c}{\textbf{\textit{Zero-shot/Few-shot Models}}} \\

  \textbf{CoT \cite{lu2023chameleon}} & GPT-4 & 85.48 & 72.44 & 90.27 & 82.65 & 71.49 & 92.89 & 86.66 & 79.04 & 83.99 \\

  \textbf{Chameleon \cite{lu2023chameleon}} & ChatGPT & 81.62 & 70.64 & 84.00 & 79.77 & 70.80 & 86.62 & 81.86 & 76.53 & 79.93 \\

  \textbf{Qwen2.5-VL-7B \cite{bai2025qwen25vl}} & Qwen2.5-VL-7B & 88.77 & 88.86 & 83.82 & 87.39 & 88.05 & 86.06 & 89.94 & 83.12 & 87.50 \\
  \textbf{Gemini-3.1-Pro \cite{gemini3.1pro}} & Gemini-3.1-Pro & 93.53 & 94.12 & 90.81 & 92.98 & 94.25 & 90.27 & 94.03 & 90.93 & 92.90 \\
  \textbf{GPT-5.2 \cite{gpt5.2}} & GPT-5.2 & 97.60 & 93.05 & 95.96 & 97.73 & 95.35 & 96.17 & 96.54 & 95.88 & 96.30 \\
  \textbf{Qwen3.5-27B \cite{qwen3.5}} & Qwen3.5-27B & \underline{98.27} & \underline{95.16} & \underline{97.18} & 97.90 & \textbf{96.43} & \underline{97.49} & \underline{97.76} & \underline{96.57} & \underline{97.34} \\
  \midrule
  \rowcolor[HTML]{FFF2CC} \multicolumn{11}{c}{\textbf{\textit{Graph Retrieval-Augmented Models}}}\\

  \textbf{MMKG \cite{liu2019mmkg}} & Qwen2.5-7B & 73.98 & 66.37 & 78.18 & 71.65 & 64.30 & 79.65 & 76.51 & 68.03 & 73.47 \\
  \textbf{Visual Genome \cite{krishna2017visual}} & Qwen2.5-7B & 76.78 & 67.04 & 78.09 & 74.05 & 66.19 & 79.72 & 78.08 & 69.68 & 75.08 \\
  \textbf{VaLiK \cite{liu2025aligning}} & Qwen2.5-7B & 84.15 & 75.14 & 87.64 & 82.99 & 73.18 & 89.69 & 84.40 & 80.95 & 83.16 \\
  \textbf{VaLiK \cite{liu2025aligning}} & Qwen2.5-72B & 85.61 & 75.93 & 90.27 & 84.40 & 74.17 & 92.33 & 85.79 & 82.98 & 84.77 \\
  \textbf{MMGraphRAG \cite{wan2025mmgraphrag}} & Qwen2.5-VL-7B & 81.08 & 68.62 & 80.09 & 79.52 & 68.96 & 83.69 & 80.87 & 73.43 & 78.21 \\
  \textbf{{MG}$^2$-RAG(Ours)} & Qwen2.5-7B & 85.17 & 81.33 & 85.82 & 83.43 & 76.10 & 89.41 & 86.12 & 81.67 & 84.53 \\
  \textbf{{MG}$^2$-RAG(Ours)} & Qwen2.5-72B & 88.37 & 81.21 & 91.27 & 87.00 & 78.28 & 94.01 & 88.84 & 85.43 & 87.62 \\
  \textbf{{MG}$^2$-RAG(Ours)} & Qwen2.5-VL-7B & 90.81 & 87.96 & 87.09 & 89.93 & 86.61 & 88.92 & 90.64 & 86.75 & 89.25 \\
  \textbf{{MG}$^2$-RAG(Ours)} & GPT-5.2 & 97.78 & 94.65 & 97.06 & \underline{97.93} & \underline{96.24} & 97.05 & 97.33 & 96.43 & 97.00 \\
  \textbf{{MG}$^2$-RAG(Ours)} & Qwen3.5-27B & \textbf{98.45} & \textbf{95.28} & \textbf{98.73} & \textbf{98.44} & \textbf{96.43} & \textbf{98.68} & \textbf{98.42} & \textbf{96.84} & \textbf{97.85} \\
  \bottomrule
\end{tabular}}
\end{table*}

Table \ref{tab:scienceqa} reports results on ScienceQA, a widely used benchmark for evaluating multimodal reasoning across diverse scientific domains. 
MG$^2$-RAG achieves state-of-the-art performance, consistently surpassing both zero-shot/few-shot MLLMs and existing graph-based RAG models. 
In particular, when paired with the Qwen3.5-27B backbone, MG$^2$-RAG reaches an overall accuracy of \textbf{97.85}, substantially exceeding human performance (88.40) \cite{lu2022learn}.

Two observations highlight the effectiveness of our framework.
First, MG$^2$-RAG consistently outperforms prior graph-augmented models.
Using the same Qwen2.5-VL-7B backbone, it improves over MMGraphRAG by 11.04 points in average accuracy and surpasses VaLiK by 1.37 points (7B) and 2.85 points (72B). 
These results demonstrate that our \textbf{multi-granularity graph representation} more effectively captures reasoning chains and structural relationships than conventional graph formulations.
Second, MG$^2$-RAG improves the reasoning capability of strong base MLLMs.
Compared with zero-shot inference, integrating MG$^2$-RAG yields average improvements of 1.75 points for Qwen2.5-VL-7B and 0.70 points for GPT-5.2.

The gains on ScienceQA further show the robustness of our \textbf{hierarchical MMKG}: when fine-grained grounding is imperfect, chunk and image nodes still provide coarser contextual evidence for reasoning.

\subsection{Multimodal Classification Performance}
\label{sec:expt:crisismmd}

\begin{table}[!t]
\centering
\small 
\setlength{\tabcolsep}{5pt} 
\renewcommand{\arraystretch}{1.1} 
\caption{\textbf{Multimodal classification performance on CrisisMMD.} Evaluation includes three tasks: \textbf{BC} (binary informativeness classification), \textbf{MC} (fine-grained humanitarian category classification), and \textbf{MC-m} (merged-category classification).}
\vspace{-0.5em}
\label{tab:crisismmd}
\resizebox{0.75\linewidth}{!}{
\begin{tabular}{ll rrrr}
  \toprule
  \multirow{2}{*}[-0.5ex]{\textbf{Method}} & \multirow{2}{*}[-0.5ex]{\textbf{LLM/MLLM}} & \multicolumn{4}{c}{\textbf{CrisisMMD}} \\
  \cmidrule(lr){3-6} 
  & & \textbf{BC$\uparrow$} & \textbf{MC$\uparrow$} & \textbf{MC-m$\uparrow$} & \textbf{Avg.$\uparrow$} \\
  \midrule

  \rowcolor[HTML]{D9EAD3} 
  \multicolumn{6}{c}{\textbf{\textit{Zero-Shot Models}}} \\
  \textbf{CLIP ViT-L/14 \cite{radford2021learning}} & CLIP ViT-L/14 & 43.36 & 17.88 & 20.79 & 27.34 \\
  \textbf{LLaVA-34B \cite{liu2023visual}} & LLaVA-34B & 56.44 & 25.15 & 25.07 & 35.55 \\
  \textbf{BLIP-2 \cite{li2023blip}} & Flan-T5 & 61.29 & 40.86 & 40.72 & 47.62 \\
   \textbf{Qwen2.5-VL-7B \cite{bai2025qwen25vl}} & Qwen2.5-VL-7B & 70.45 & 41.44 & 42.33 & 51.41 \\
  \textbf{Gemini-3.1-Pro \cite{gemini3.1pro}} & Gemini-3.1-Pro & 71.30 & 49.36 & 49.79 & 56.82 \\
  \textbf{GPT-5.2 \cite{gpt5.2}} & GPT-5.2 & 71.00 & \underline{53.00} & \underline{53.60} & \underline{59.20} \\
  \textbf{Qwen3.5-27B \cite{qwen3.5}} &Qwen3.5-27B & 69.83 & 48.82 & 49.35 & 56.00 \\
  \midrule

  \rowcolor[HTML]{FFF2CC} \multicolumn{6}{c}{\textbf{\textit{Graph Retrieval-Augmented Models}}} \\
  \textbf{LightRAG \cite{guo2024lightrag}} & Qwen2.5-7B & 67.49 & 45.11 & 45.94 & 52.85 \\
  \textbf{VaLiK \cite{liu2025aligning}} & Qwen2.5-7B & 68.90 & 50.02 & 50.69 & 56.54 \\
  \textbf{VaLiK \cite{liu2025aligning}} & Qwen2.5-72B & 68.89 & 49.78 & 49.31 & 55.99 \\
  \textbf{MMGraphRAG \cite{wan2025mmgraphrag}} & Qwen2.5-VL-7B & 68.40 & 44.03 & 44.66 & 50.94 \\
  \textbf{{MG}$^2$-RAG(Ours)} & Qwen2.5-VL-7B & 71.03 & 46.45 & 47.25 & 54.91 \\
  \textbf{{MG}$^2$-RAG(Ours)} & Qwen2.5-7B & 69.69 & 51.1 & 51.59 & 57.46 \\
  \textbf{{MG}$^2$-RAG(Ours)} & Qwen2.5-72B & \textbf{72.33} & 50.34 & 50.83 & 57.83 \\
  \textbf{{MG}$^2$-RAG(Ours)} & Qwen3.5-27B & 72.28 & 52.30 & 52.79 & 59.12 \\
  \textbf{{MG}$^2$-RAG(Ours)} & GPT-5.2 & \underline{72.30} & \textbf{56.10} & \textbf{56.70} & \textbf{61.70} \\
  \bottomrule
\end{tabular}}
\end{table}

Table \ref{tab:crisismmd} presents results on CrisisMMD, evaluating multimodal classification in disaster scenarios. 
MG$^2$-RAG achieves state-of-the-art performance. 
In particular, the combination of MG$^2$-RAG and GPT-5.2 obtains the best overall accuracy of \textbf{61.70}, demonstrating strong performance in both binary relevance filtering (BC) and multi-class humanitarian classification (MC and MC-m).

Two insights emerge from the results.
First, MG$^2$-RAG consistently surpasses existing graph-based retrieval frameworks. 
Using the Qwen2.5-VL-7B backbone, our model improves over MMGraphRAG by 3.97 points in average accuracy. 
With Qwen2.5-7B, it outperforms LightRAG by 4.61 points and VaLiK by 0.92 points, while maintaining a 1.84-point advantage over VaLiK with the 72B backbone. 
These improvements highlight the effectiveness of our \textbf{multi-granularity graph retrieval mechanism}, which captures complex cross-modal relationships within noisy real-world data.
Second, MG$^2$-RAG substantially enhances the zero-shot capability of standard MLLMs. 
Compared with their respective zero-shot baselines, our framework improves performance by 3.50 points for Qwen2.5-VL-7B and 2.50 points for GPT-5.2. 
This demonstrates that our \textbf{entity-driven visual grounding} helps resolve ambiguity in social-media content and improves decision reliability.

\subsection{Ablation Study}
\label{sec:expt:ablation}

\begin{table}[t]
\centering
\small
\setlength{\tabcolsep}{6pt} 
\renewcommand{\arraystretch}{1.0}
\caption{\textbf{Ablation study of MG$^2$-RAG.}
HGC, MNF, and GP denote Hierarchical Graph Construction, Multimodal Node Fusion, and Graph Propagation, respectively.}
\vspace{-0.5em}
\label{tab:camera_ablation}
\resizebox{0.85\columnwidth}{!}{%
\begin{tabular}{lcccc}
  \toprule
  \multirow{2}{*}{\textbf{Method}} &
  \textbf{E-VQA (5k)} &
  \textbf{E-VQA (5k)} &
  \textbf{ScienceQA} &
  \textbf{CrisisMMD} \\
  & \textbf{R@1/R@5$\uparrow$} &
  \textbf{BEM Score (All$\uparrow$)} &
  \textbf{Acc. (Avg.$\uparrow$)} &
  \textbf{Acc. (Avg.$\uparrow$)} \\
  \midrule
  \textbf{MG$^2$-RAG} & \textbf{57.8 / 83.1} & \textbf{60.24} & \textbf{97.85} & \textbf{59.12} \\
  \quad w/o HGC & 40.7 / 61.5 & 51.72 & 97.48 & 56.70 \\
  \quad w/o MNF & 43.8 / 78.0 & 55.38 & 97.67 & 58.62 \\
  \quad w/o GP & 25.5 / 76.0 & 54.69 & 97.51 & 58.90 \\
  \bottomrule
\end{tabular}}
\end{table}

Table~\ref{tab:camera_ablation} presents an ablation study of the three key components in MG$^2$-RAG: Hierarchical Graph Construction (\textbf{HGC}), Multimodal Node Fusion (\textbf{MNF}), and Graph Propagation (\textbf{GP}).
Removing any component consistently degrades performance, demonstrating that the three designs complement each other to enable effective multimodal retrieval.
Two observations are particularly noteworthy.

First, the \textbf{hierarchical graph structure} is essential for organizing multimodal evidence.
Removing HGC collapses the hierarchical chunk-image-node representation into a flatter retrieval space, reducing E-VQA R@1/R@5 from 57.8/83.1 to 40.7/61.5. This confirms that the hierarchy effectively bridges coarse document context with fine-grained visual and textual evidence.
Second, \textbf{cross-modal node alignment} and \textbf{graph-based relevance propagation} improve retrieval from complementary perspectives.
Removing MNF decreases the E-VQA BEM score from 60.24 to 55.38, demonstrating the importance of aligning textual entities with grounded visual regions. In contrast, removing GP causes the largest drop in retrieval accuracy, reducing R@1 to 25.5, indicating that graph propagation is crucial for discovering structurally related evidence beyond initial dense retrieval.
Together, these results show that MG$^2$-RAG benefits from hierarchy, multimodal alignment, and graph propagation in a complementary way.

\subsection{Case Study}
\label{sec:expt:qualitative_results}

\begin{figure*}[t]
  \centering
  \includegraphics[width=0.99\textwidth]{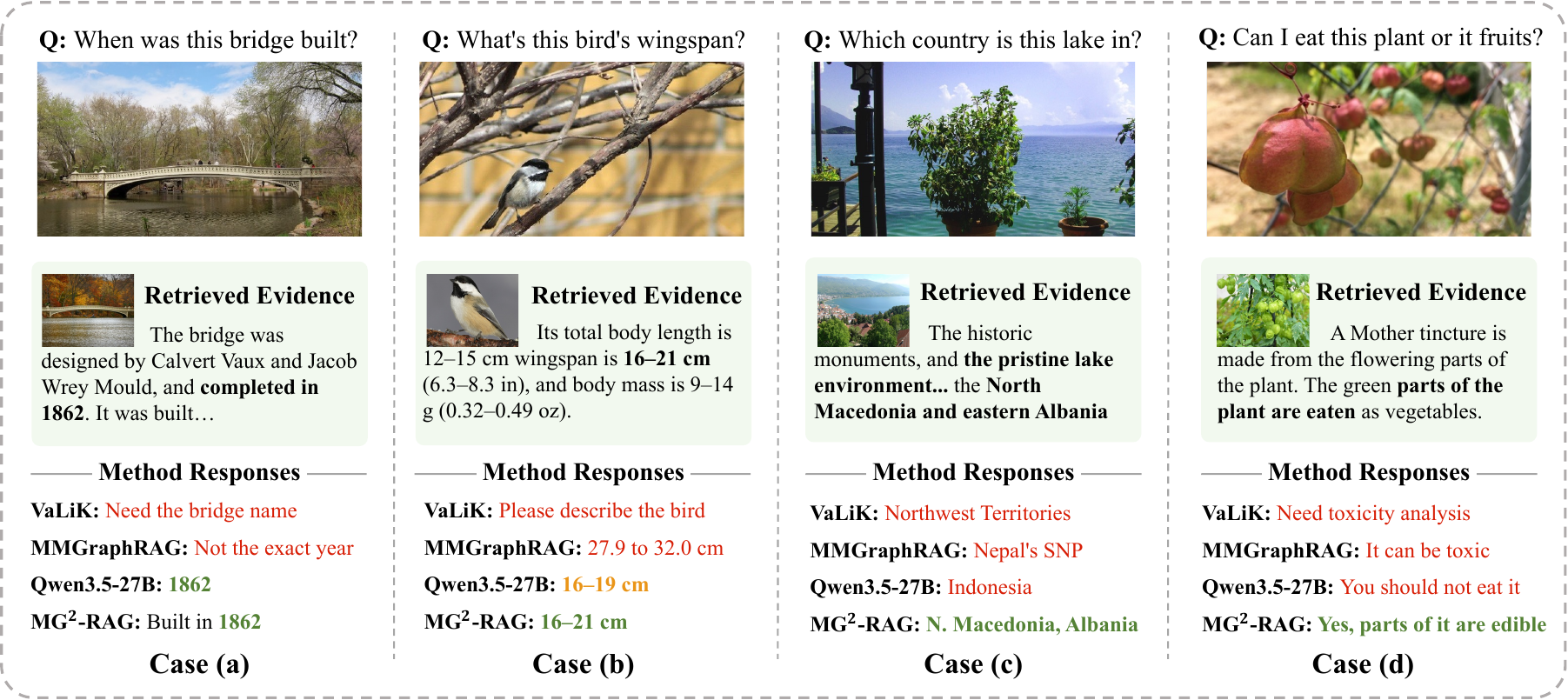}
  \vspace{-0.5em}
  \caption{\textbf{Case studies on Knowledge-based VQA and Multimodal Retrieval.} MG$^2$-RAG is compared with a baseline MLLM (Qwen3.5-27B) and graph-based approaches (VaLiK and MMGraphRAG). The \textbf{Retrieved Evidence} illustrates the multimodal evidence retrieved by MG$^2$-RAG for answer generation.}
  \label{fig:vqa-casestudy}
  \vspace{-0.5em}
\end{figure*}

Figure \ref{fig:vqa-casestudy} presents qualitative comparisons on knowledge-based VQA and multimodal retrieval.
Baseline MLLMs (e.g., Qwen3.5-27B) frequently generate hallucinated or inaccurate answers due to the lack of external knowledge. 
For example, in Case (c), the baseline incorrectly predicts \textit{Indonesia}. In contrast, MG$^2$-RAG produces accurate, grounded responses by \textbf{fusing textual entities and visual objects into unified multimodal nodes}, preserving fine-grained cross-modal evidence. This enables precise retrieval of details such as the wingspan in Case (b) and plant edibility in Case (d).
Moreover, \textbf{multi-granularity graph retrieval} supports effective cross-modal reasoning.
In Case (c), MG$^2$-RAG retrieves the key evidence ``the pristine lake environment'' and propagates it to the corresponding image, correctly identifying the location as ``North Macedonia, Albania.''
Additional qualitative examples for multimodal reasoning, classification, and representative failure cases are provided in Appendix \ref{app:case-study}.

\section{Conclusion}
\label{sec:conclusion}

In this paper, we introduce \textbf{MG$^2$-RAG}, a lightweight Multi-Granularity Graph RAG framework that improves the reliability and efficiency of MM-RAG. 
We propose a hierarchical MMKG construction strategy that bypasses expensive MLLM-driven triplet extraction by combining lightweight textual parsing with entity-driven visual grounding, enabling efficient fusion of textual entities and visual objects into unified multimodal nodes. 
Building on this representation, we further design a multi-granularity graph retrieval mechanism that aggregates dense similarities and propagates relevance through the graph topology to support structured multi-hop reasoning.
Extensive experiments across four multimodal tasks (i.e., retrieval, knowledge-based VQA, reasoning, and classification) show that MG$^2$-RAG consistently achieves state-of-the-art performance while maintaining high efficiency, providing an average $43.3\times$ speedup and $23.9\times$ cost reduction compared with existing graph-based baselines.
Ultimately, MG$^2$-RAG provides a scalable, efficient, and robust solution for MM-RAG, enabling reliable deployment of MLLMs in complex, real-world knowledge-intensive applications.

\section*{Acknowledgements}
This work was supported by the National Natural Science Foundation of China (NSFC) under Grant Nos. 62125201, U24B20174, and U25B6003, as well as the New Generation Artificial Intelligence-National Science and Technology Major Project (2025ZD0123302).

\bibliographystyle{splncs04}
\bibliography{main}
\newpage
\appendix
\section{Implementation Details}
\label{app:impl}

\subsection{Rule-based Relation Extraction} 
\label{app:impl:relation-extraction}

To construct semantic relations between named entities during textual graph construction, we employ a lightweight rule-based extraction strategy built on the dependency parsing trees generated by \texttt{spaCy} \cite{honnibal2020spacy}. 
Unlike prior graph-based RAG systems that rely on expensive LLM-driven triplet extraction, our approach directly leverages grammatical structures to identify high-precision relations while maintaining computational efficiency, consistent with the lightweight Multimodal Knowledge Graph (MMKG) construction pipeline described in the main paper.

The extraction procedure operates \emph{sentence-by-sentence}. 
For each sentence, we first identify all named entities and retain only \textbf{relational entities} $\mathcal{E}$ whose labels exclude \texttt{ORDINAL}, \texttt{CARDINAL}, \texttt{PERCENT}, or \texttt{QUANTITY}.
When at least two relational entities are present ($|\mathcal{E}| \geq 2$), we examine each unique entity pair $(e_1, e_2)$ using \textbf{dependency-root alignment and part-of-speech (POS) analysis}.
To capture the dominant grammatical patterns in natural language, the extraction rules are organized into two complementary categories: 
\begin{itemize}
  \item \textbf{Verb-Centric Structures}, which capture relations expressed via predicates.
  
  \item \textbf{Noun-Centric Structures}, which extract relations from nominal constructions when no governing verb is present.
\end{itemize}
For consistency across syntactic variations, all relation predicates are normalized using the \emph{lemma form of the governing token}.

\paragraph{Verb-Centric Rules} 
Verb-centric rules handle sentences where the relation between two entities is mediated by a verbal predicate.
When the dependency roots of $e_1$ and $e_2$ share the same governing verb, or are connected through a verb-preposition construction, the relation is derived from the lemma of the verb.
\begin{itemize}
  \item \textbf{Predicate-Centered Structures:} 
  If both entities attach to the same verbal root as a \textbf{subject} (\texttt{nsubj} or \texttt{csubj}) and \textbf{object} (\texttt{dobj} or \texttt{attr}), the relation is extracted directly from the verb.
  If the verb contains a negation modifier (\texttt{neg}), the predicate is prefixed with \emph{not}.
  The resulting triplet is denoted as:
  \[
    \langle e_1, [\text{not}] + \text{verb}_{lemma}, e_2 \rangle.
  \]
  This rule captures the most common \emph{active predicate constructions}.

  \item \textbf{Adpositional Structures:} 
  When the dependency path between two entities includes an \textbf{adposition} (\texttt{ADP}), the relation predicate is expanded to incorporate the preposition.
  For standard active patterns, we append the preposition to the verb lemma: 
  \[
    \langle e_1, \text{verb} + \text{prep}, e_2 \rangle.
  \]
  For \textbf{passive constructions}, we apply a passive correction mechanism.
  If the dependency pattern indicates a passive subject (\texttt{nsubjpass}) and the preposition \emph{by}, the entity roles are inverted to reconstruct the corresponding active relation.
  This normalization ensures consistent relational representations regardless of syntactic voice.
  The rules also handle inverted prepositional structures detected through symmetric dependency patterns.
\end{itemize}

\begin{figure*}[t]
  \centering
  \includegraphics[width=0.99\textwidth]{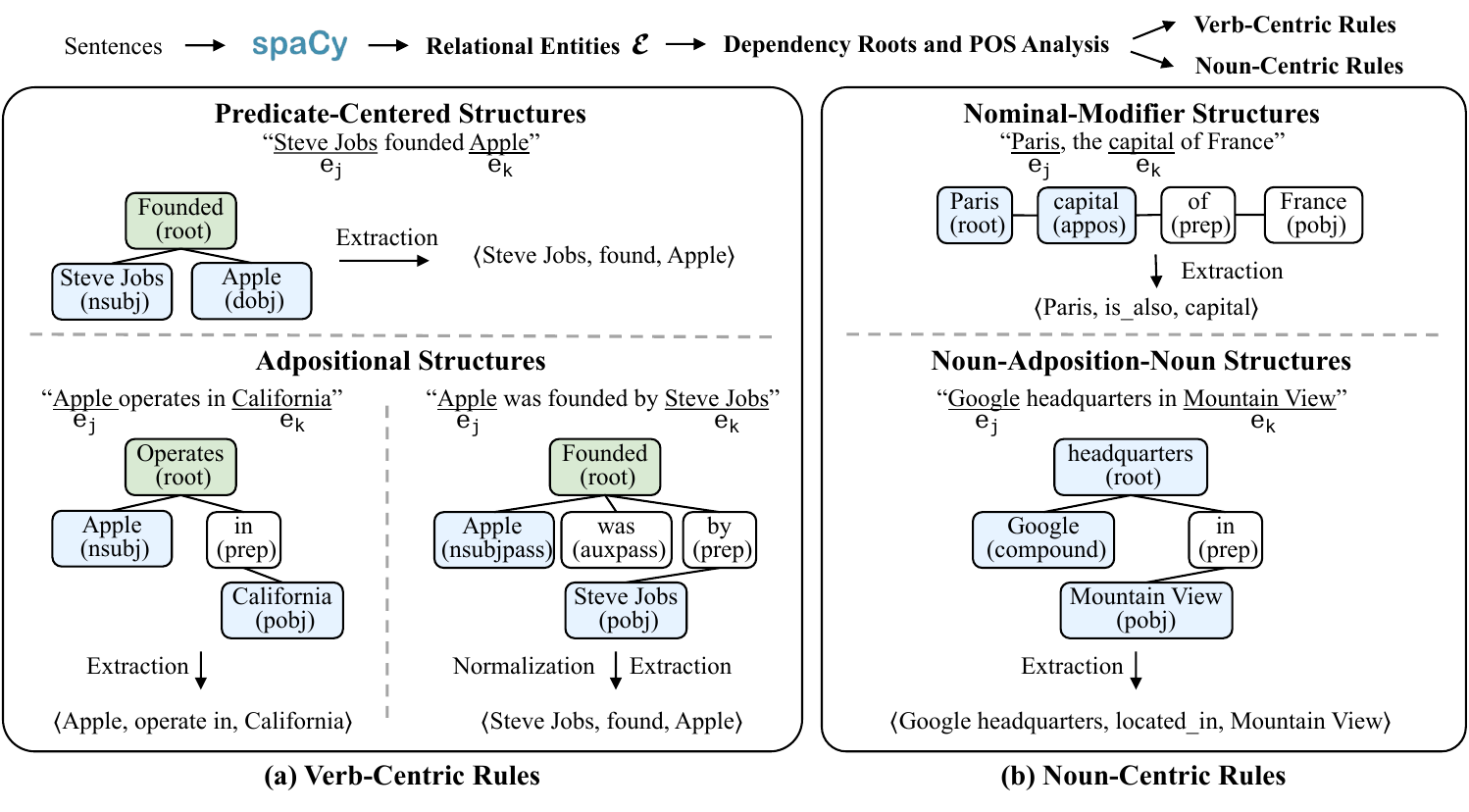}
  \vspace{-0.5em}
  \caption{\textbf{Rule-based Relation Extraction Examples.} The extraction framework is divided into two syntactic categories: \textbf{(a) Verb-Centric Rules}, which capture relations expressed through predicates and verb–preposition structures, and \textbf{(b) Noun-Centric Rules}, which derive relations from nominal modifiers and noun-preposition constructions when no governing verb is present.}
  \label{fig:relation-extraction}
  \vspace{-0.5em}
\end{figure*}

\paragraph{Noun-Centric Rules} 
When no governing verb connects two entities, we extract relations from nominal modifiers and adpositional attachments.
\begin{itemize}
  \item \textbf{Nominal-Modifier Structures:} 
  If two entities are linked through an appositional modifier (\texttt{appos}), the construction indicates semantic equivalence or renaming.
  In such cases, we extract the relation:
  \[
    \langle e_1, \text{is\_also}, e_2 \rangle,
  \]
  which represents one entity functions as an alternative description of the other.

  \item \textbf{Noun-Adposition-Noun Structures:} 
  When two entities are connected via an adpositional phrase without an intervening verb, the adposition itself carries the relational meaning.
  In this case, the preposition is mapped to a predefined semantic predicate.
  For example, spatial prepositions like \emph{in} or \emph{at} are normalized to the relation \texttt{located\_in}, producing triplets of the form
  \[
    \langle e_1, \text{located\_in}, e_2 \rangle.
  \]
\end{itemize}

\paragraph{Extraction Examples} 
Figure~\ref{fig:relation-extraction} illustrates several representative cases demonstrating the application of these rules.
For \textbf{verb-centric structures}, the sentence \emph{``Steve Jobs founded Apple''} contains a subject–verb–object pattern, yielding the triplet $\langle \text{Steve Jobs}, \text{found}, \text{Apple} \rangle$.
Similarly, the sentence \emph{``Apple operates in California''} contains a verb–preposition construction, resulting in $\langle \text{Apple}, \text{operate\_in}, \text{California} \rangle$.
For passive constructions such as \emph{``Apple was founded by Steve Jobs,''} the passive correction rule reverses the entity order, producing the same active relation $\langle \text{Steve Jobs}, \text{found}, \text{Apple} \rangle$.

For \textbf{noun-centric structures}, the appositive phrase \emph{``Paris, the capital of France''} triggers the nominal modifier rule, yielding
$\langle \text{Paris}, \text{is\_also}, \text{capital} \rangle$.
Finally, in the phrase \emph{``Google headquarters in Mountain View,''} the adpositional relation is mapped to a spatial predicate, producing $\langle \text{Google headquarters}$, $\text{located\_in}, \text{Mountain View} \rangle$.

\subsection{Dataset Pre-Processing} 
\label{app:impl:dataset}

\paragraph{Dataset Split Strategies}
For the Multimodal Retrieval and knowledge-based Visual Question Answering (VQA) tasks, we evaluate our approach on two large-scale benchmarks: \textbf{E-VQA} \cite{mensink2023encyclopedic} and \textbf{InfoSeek} \cite{chen2023can}.
To maintain a consistent retrieval scale across our experiments, we follow standard practices established for InfoSeek \cite{yuan2025mkg, cocchi2025augmenting, yan2024echosight} and restrict the external knowledge base for both datasets to 100k documents. 
The specific splitting and pre-processing details for each dataset are as follows:
\begin{itemize}
  \item \textbf{E-VQA \cite{mensink2023encyclopedic}:} 
  This dataset comprises 221k question-answer pairs, each paired with up to 5 images and associated with 16.7k fine-grained entities. 
  Questions are categorized as either single-hop or two-hop. 
  Single-hop queries require a single Wikipedia page to answer, whereas two-hop queries necessitate sequential retrieval across multiple documents. 
  The dataset is partitioned into training, validation, and test splits, containing 1M, 13.6k, and 5.8k samples, respectively. 
  All our experiments are conducted on the test split, which includes 4.8k single-hop questions. The provided external knowledge base, derived from Wikipedia, contains approximately 2M pages. 
  As mentioned above, we randomly sample a subset of 100k documents from this original knowledge base for our evaluation.

  \item \textbf{InfoSeek \cite{chen2023can}:} 
  The InfoSeek dataset consists of approximately 1.3M image-question-answer triplets corresponding to around 11k distinct Wikipedia pages. 
  It is divided into training, validation, and test splits, containing roughly 934k, 73k, and 348k samples, respectively. 
  Notably, both the validation and test sets feature questions regarding unseen entities. 
  InfoSeek provides an external knowledge base of around 6M Wikipedia entities. 
  we follow previous works \cite{yuan2025mkg, cocchi2025augmenting, yan2024echosight} and conduct our experiments using a subset of 100k pages.
\end{itemize}

\paragraph{Baseline Comparison Settings}
For the multimodal retrieval and knowledge-based VQA tasks, to facilitate comparisons with baseline models, we construct smaller dataset variants from the original benchmarks. 
Since baseline approaches \cite{liu2025aligning, wan2025mmgraphrag} rely on expensive MLLM-driven triplet extraction, making full-scale graph construction computationally and financially infeasible, these smaller variants are necessary for evaluation. 

The detailed splitting and pre-processing strategies are as follows:
For the E-VQA and InfoSeek benchmarks, we restrict the knowledge base to a 5k subset. 
For this restricted 5k subset, we ensure the inclusion of the necessary evidence required to answer the evaluation questions. 
To align the evaluation scales, the E-VQA test set remains unchanged at 5.8k samples, while we randomly extract 5.8k samples from the InfoSeek 73k validation set.

\paragraph{Closed-Source Models Evaluation} 
For the evaluation of closed-source MLLMs \cite{gemini3.1pro, gpt5.2}, we evaluate on the E-VQA \cite{mensink2023encyclopedic}, InfoSeek \cite{chen2023can}, ScienceQA \cite{lu2022learn}, and CrisisMMD  \cite{alam2018crisismmd} (including both binary informativeness classification and fine-grained humanitarian category classification) datasets. 
Rather than using any restricted subsets, we randomly sample 1k instances directly from the original evaluation splits of each dataset (e.g., the E-VQA test set and the InfoSeek validation set).

\paragraph{Efficiency and Cost Evaluation}
To ensure a fair comparison regarding time efficiency, both our proposed method and the baselines are evaluated on the same hardware infrastructure. 
For the cost evaluation, we specifically focus on the expenses incurred during the knowledge graph construction phase. 
The baseline approaches rely on LLMs to construct the graph; therefore, their cost is measured by the token consumption based on public API pricing (i.e., Qwen-2.5-VL-7B-Instruct via OpenRouter\footnote{\url{https://openrouter.ai/qwen/qwen-2.5-vl-7b-instruct}.}).
In contrast, our graph construction process is completely token-free. 
Consequently, our cost is solely determined by the GPU rental expenses accumulated during the processing time, calculated using standard rates from the AutoDL platform.\footnote{\url{https://www.autodl.com/market/list}}

\subsection{Hyperparameter Configurations and Prompts of MG$^2$-RAG}
\label{app:iml:task-setting}

\paragraph{Multimodal Retrieval Task}
As a comprehensive Multimodal Retrieval-Augmented Generation (MM-RAG) framework, MG$^2$-RAG fundamentally relies on its retrieval component to support all downstream reasoning tasks.
In this stage, the framework performs multi-granularity retrieval over the constructed multimodal knowledge graph to identify fine-grained textual and visual evidence relevant to the query.
To provide a detailed overview of the retrieval configurations across different datasets, we summarize the corresponding hyperparameter settings in Table~\ref{tab:hyperparameters}. 

\begin{table}[t]
\centering
\small 
\caption{\textbf{MG$^2$-RAG hyperparameter configurations across different datasets.} These parameters are specifically employed during the Multi-Granularity Graph Retrieval process to accommodate the distinct characteristics of various tasks.}
\label{tab:hyperparameters}
\vspace{-0.5em}
\setlength{\tabcolsep}{5pt} 
\renewcommand{\arraystretch}{1.2} 
\resizebox{\linewidth}{!}{%
\begin{tabular}{l ccccc}
  \toprule
  \textbf{Parameter} & \textbf{E-VQA} & \textbf{Infoseek} & \textbf{ScienceQA} & \textbf{Task-BC} & \textbf{Task-MC} \\
  \midrule
  \rowcolor[HTML]{D9EAD3}
  \multicolumn{6}{c}{\textbf{\textit{Graph Propagation and Modality Fusion}}} \\
  Damping factor ($\alpha$) & $0.20$ & $0.15$ & $0.85$ & $0.85$ & $0.70$ \\
  Chunk Node Weight ($\omega_C$) & $0.8$ & $1.2$ & $0.05$ & $0.2$ & $1.0$ \\
  Image Node Weight ($\omega_I$) & $1.6$ & $0.5$ & $1.0$ & $1.0$ & $1.0$ \\
  Textual Fusion Weight ($\lambda_t$) & $0.1$ & $0.1$ & $1.0$ & $1.0$ & $1.0$ \\
  Visual Fusion weight ($\lambda_v$) & $1.0$ & $1.0$ & $1.0$ & $0.5$ & $1.0$ \\
  \midrule
  \rowcolor[HTML]{E7F3FF} 
  \multicolumn{6}{c}{\textbf{\textit{Activated Seed Nodes by Textual Query}}} \\
  Chunk top-$k$ & $60$ & $200$ & $4$ & $7$ & $12$ \\
  Sentence top-$k$ & $3$ & $3$ & $10$ & $5$ & $2$ \\
  Image top-$k$ & $2$ & $2$ & $10$ & $5$ & $3$ \\
  Object top-$k$ & $3$ & $3$ & $10$ & $5$ & $2$ \\
  \midrule
  \rowcolor[HTML]{FFF2CC} 
  \multicolumn{6}{c}{\textbf{\textit{Activated Seed Nodes by Visual Query}}} \\
  Chunk top-$k$ & $200$ & $200$ & $10$ & $3$ & $12$ \\
  Sentence top-$k$ & $70$ & $60$ & $10$ & $1$ & $3$ \\
  Image top-$k$ & $2$ & $1$ & $20$ & $1$ & $5$ \\
  Object top-$k$ & $5$ & $5$ & $10$ & $1$ & $3$ \\
  \bottomrule
\end{tabular}}
\end{table}
\setlength{\textfloatsep}{1.5em}

These parameters are adapted to the characteristics of each task to ensure effective retrieval.
For instance, in VQA, textual queries such as ``what is this'' often contain limited semantic information and therefore contribute less to retrieval.
To address this issue, we adjust modality weights and seed node activations to emphasize visual evidence.

Despite these task-specific configurations, several parameters remain consistent across all settings.
In particular, we fix the visual grounding threshold at $\tau = 0.5$ and set the Personalized PageRank \cite{haveliwala2002topic} convergence tolerance to $\epsilon = 10^{-6}$. 
The multimodal evidence retrieved in this stage serves as the grounding context for the generation process, providing structured information that is subsequently integrated into the prompts for the VQA, reasoning, and classification tasks described below.

To evaluate the robustness of MG$^2$-RAG to retrieval-stage hyperparameters, we conduct a series of controlled experiments on E-VQA (5k), using R@1 as the evaluation metric.
Specifically, we analyze the effects of modality fusion weights ($\lambda_v,\lambda_t$), node scaling factors ($\omega_I,\omega_C$), the number of top-$k$ chunk seeds, and the graph propagation coefficient ($\alpha$). The corresponding results are summarized in Tables~\ref{tab:camera_parameter_modality}--\ref{tab:camera_parameter_propagation}.

MG$^2$-RAG exhibits stable performance across a wide range of hyperparameter settings. 
The highest or near-highest R@1 is consistently achieved around the default configuration adopted in the main experiments, while moderate variations in the retrieval hyperparameters lead to only marginal performance changes. 
These results demonstrate that the effectiveness of MG$^2$-RAG primarily stems from its hierarchical graph representation and multi-granularity retrieval mechanism, rather than careful hyperparameter tuning.

\paragraph{Knowledge-based VQA Task}
For Knowledge-based VQA, the hyperparameter configurations emphasize visual evidence during retrieval. 
Both E-VQA and InfoSeek assign a high visual fusion weight ($\lambda_v = 1.0$) and a low textual fusion weight ($\lambda_t = 0.1$), reflecting the dominant role of visual queries.
In addition, both datasets use a low damping factor ($\alpha \le 0.20$), favoring the retrieval of highly relevant local evidence rather than deep multi-hop propagation.

Despite these similarities, the datasets require different node-weighting strategies. 
E-VQA assigns a higher weight to image nodes ($\omega_I = 1.6$) to support visual-centric queries, whereas InfoSeek prioritizes chunk nodes ($\omega_C = 1.2$) to retrieve external textual knowledge grounded in the visual input.
The retrieved evidence is then combined with the original user query to construct prompts for downstream models. We adopt task-specific prompt templates, following EchoSight \cite{yan2024echosight} for E-VQA and OMGM \cite{yang-etal-2025-omgm} for InfoSeek.

\begin{table}[!t]
\centering
\small
\setlength{\tabcolsep}{6pt}
\renewcommand{\arraystretch}{1.0}
\caption{\textbf{Sensitivity to modality weights on E-VQA (5k).}
R@1 remains stable under moderate changes of visual and textual fusion weights.}
\vspace{-0.5em}
\label{tab:camera_parameter_modality}
\resizebox{0.8\columnwidth}{!}{%
\begin{tabular}{lccccc ccccc}
  \toprule
  \multirow{2}{*}{\textbf{Weight}} &
  \multicolumn{5}{c}{$\bm{\lambda_v}$} &
  \multicolumn{5}{c}{$\bm{\lambda_t}$} \\
  \cmidrule(lr){2-6} \cmidrule(lr){7-11}
  & 0.5 & 0.8 & 1.0 & 1.2 & 1.5 & 0.05 & 0.08 & 0.10 & 0.12 & 0.15 \\
  \midrule
  \textbf{R@1$\uparrow$} & 56.7 & \textbf{57.8} & \textbf{57.8} & 57.7 & 57.7 & 57.6 & 57.7 & \textbf{57.8} & 57.7 & 57.6 \\
  \bottomrule
\end{tabular}}
\end{table}

\begin{table}[!t]
\centering
\small
\setlength{\tabcolsep}{6pt}
\renewcommand{\arraystretch}{1.0}
\caption{\textbf{Sensitivity to node scaling factors on E-VQA (5k).} R@1 is stable across different image-node and chunk-node weights.}
\vspace{-0.5em}
\label{tab:camera_parameter_scaling}
\resizebox{0.9\columnwidth}{!}{%
\begin{tabular}{lccccc ccccc}
  \toprule
  \multirow{2}{*}{\textbf{Scaling Factor}} &
  \multicolumn{5}{c}{$\bm{\omega_I}$} &
  \multicolumn{5}{c}{$\bm{\omega_C}$} \\
  \cmidrule(lr){2-6} \cmidrule(lr){7-11}
  & 1.1 & 1.4 & 1.6 & 1.8 & 2.1 & 0.3 & 0.6 & 0.8 & 1.0 & 1.3 \\
  \midrule
  \textbf{R@1$\uparrow$} & 57.7 & \textbf{57.8} & \textbf{57.8} & \textbf{57.8} & 57.7 & 57.6 & \textbf{57.8} & \textbf{57.8} & 57.6 & 57.7 \\
  \bottomrule
\end{tabular}}
\end{table}

\begin{table}[!t]
\centering
\small
\setlength{\tabcolsep}{6pt}
\renewcommand{\arraystretch}{1.0}
\caption{\textbf{Sensitivity to top-$k$ chunk seed selection on E-VQA (5k).} R@1 changes only slightly across different visual-query and textual-query seed sizes.}
\vspace{-0.5em}
\label{tab:camera_parameter_topk}
\resizebox{0.9\columnwidth}{!}{%
\begin{tabular}{lccccc ccccc}
  \toprule
  \multirow{2}{*}{\textbf{Seed Selection}} &
  \multicolumn{5}{c}{\textbf{Top-$k$ (Visual Query)}} &
  \multicolumn{5}{c}{\textbf{Top-$k$ (Textual Query)}} \\
  \cmidrule(lr){2-6} \cmidrule(lr){7-11}
  & 150 & 180 & 200 & 220 & 250 & 10 & 40 & 60 & 80 & 110 \\
  \midrule
  \textbf{R@1$\uparrow$} & 57.5 & \textbf{57.8} & \textbf{57.8} & 57.6 & 57.5 & 57.7 & 57.7 & \textbf{57.8} & 57.7 & 57.7 \\
  \bottomrule
\end{tabular}}
\end{table}

\begin{table}[!t]
\centering
\small
\setlength{\tabcolsep}{6pt}
\renewcommand{\arraystretch}{1.0}
\caption{\textbf{Sensitivity to propagation strength on E-VQA (5k).} MG$^2$-RAG performs best around the default propagation setting.}
\vspace{-0.5em}
\label{tab:camera_parameter_propagation}
\resizebox{0.7\columnwidth}{!}{%
\begin{tabular}{lccccc}
  \toprule
  \textbf{Propagation Strength ($\bm{\alpha}$)} & 0.05 & 0.10 & 0.20 & 0.30 & 0.50 \\
  \midrule
  \textbf{R@1$\uparrow$} & 55.1 & 57.2 & \textbf{57.8} & 56.0 & 55.5 \\
  \bottomrule
\end{tabular}}
\end{table}

\paragraph{Multimodal Reasoning Task}
For multimodal reasoning on ScienceQA, the configuration reflects the need for multi-hop reasoning across heterogeneous evidence. 
This is indicated by a high damping factor ($\alpha = 0.85$), which encourages deeper graph propagation to connect distributed multimodal cues.
The task also requires balanced multimodal understanding, and therefore assigns equal fusion weights to textual and visual queries ($\lambda_t = 1.0, \lambda_v = 1.0$). 
After retrieval, the collected contextual evidence is integrated with the question and candidate options. 
A unified prompt template is used to guide both LLMs and MLLMs in performing step-by-step reasoning and answer selection, as shown in Prompt~\ref{prompt:reasoning_combined}.\footnote{Italicized instructions in the prompt indicate modality-specific guidance.}

\begin{promptbox}[prompt:reasoning_combined]{Multimodal Reasoning (ScienceQA)}
\textbf{[System]} \\
You are an expert science educator. Answer the multiple-choice science question.\\ \textit{[MLLM: Do not mention the visual content in your output; base your answer directly on the image and context.]} \\
\\
\textbf{[Instructions]} \\
1. Evaluate Context: Retrieved knowledge is for reference. Use it if relevant to support reasoning; ignore it if noisy and rely on your own knowledge. \\
2. Reasoning: Briefly explain your logic step-by-step. \\
3. Format: The final answer MUST be the Option Letter only (e.g., A, B, C, D, E). \\
\\
\textbf{[User Input]} \\
\textit{[MLLM: \{Query Image\}]} \\
Retrieved Context: \$\{context\_user\} \\
Question: \$\{question\_options\_user\} \\
\\
\textbf{[Response]} \\
Reasoning:
\end{promptbox}

\begin{promptbox}[prompt:task_bc_combined]{Multimodal Classification (Task-BC)}
\textbf{[System]} \\
You are an expert in crisis response and social media analysis. \\
\textit{LLM: Answer the multiple-choice classification question}\\
\textit{[MLLM: Classify the crisis-related tweet based on its text and the provided image]}. \\
\\
\textbf{[Instructions]} \\
1. Context Usage: Retrieved tweets and labels are reference criteria. Use them if relevant; otherwise, rely on your own judgment. \\
2. Reasoning: Briefly explain your logic step-by-step\\
\textit{[MLLM: explicitly considering both visual and textual evidence]}. \\
3. Format: The final answer MUST be the Option Letter only. \\
\\
\textbf{[User Input]} \\
\textit{[MLLM: \{Query Image\}]} \\
Retrieved Context: \$\{context\_user\} \\
Target Tweet: \$\{question\_user\} \\
Options: \\
{[A]} not\_informative | {[B]} informative \\
\\
\textbf{[Response]} \\
Reasoning:
\end{promptbox}
\begin{promptbox}[prompt:task_mc_combined]{Multimodal Classification (Task-MC)}
\textbf{[System]} \\
You are an expert in crisis response and humanitarian aid. \\
\textit{[LLM: Answer the multiple-choice classification question.]} \\
\textit{[MLLM: Classify the humanitarian category of the tweet based on both its text and the provided image.]} \\
\\
\textbf{[Instructions]} \\
1. Reasoning: Briefly explain your logic step-by-step\\
\textit{[MLLM: explicitly considering both visual and textual evidence]}. \\
2. Format: The final answer MUST be the Option Letter only. \\
\\
\textbf{[User Input]} \\
\textit{[MLLM: \{Query Image\}]} \\
Retrieved Context: \$\{context\_user\} \\
Target Tweet: \$\{question\_user\} \\
Options: \\
{[A]} infrastructure\_and\_utility\_damage | {[B]} not\_humanitarian \\
{[C]} other\_relevant\_information \\ {[D]} rescue\_volunteering\_or\_donation\_effort \\
{[E]} vehicle\_damage | {[F]} affected\_individuals \\
{[G]} injured\_or\_dead\_people | {[H]} missing\_or\_found\_people \\
\\
\textbf{[Response]} \\
Reasoning:
\end{promptbox}

\paragraph{Multimodal Classification Task}
The multimodal classification setting includes two sub-tasks: \textbf{Task-BC} (binary informativeness classification) and \textbf{Task-MC} (multi-class fine-grained humanitarian category classification). 
Both tasks benefit from broader contextual aggregation, which is reflected in relatively high damping factors ($\alpha = 0.85$ for Task-BC and $\alpha = 0.70$ for Task-MC).

Unlike VQA tasks, social-media posts contain rich textual semantics. Consequently, both tasks adopt a maximum textual fusion weight ($\lambda_t = 1.0$).
The visual fusion weight is then adjusted according to the classification objective.
Task-BC uses a moderate visual weight ($\lambda_v = 0.5$) because determining informativeness often relies primarily on textual cues.
In contrast, Task-MC sets $\lambda_v = 1.0$ to capture fine-grained visual signals, such as distinguishing infrastructure damage from rescue operations, that are essential for accurate multi-class classification.

After retrieval, the contextual evidence (e.g., related tweets and historical labels) is formatted as structured references to guide the classification process. The prompt templates used for this stage are provided in Prompts~\ref{prompt:task_bc_combined} and \ref{prompt:task_mc_combined}.

\section{Additional Case Study Examples}
\label{app:case-study}

\begin{figure*}[t]
  \centering
  \includegraphics[width=0.99\textwidth]{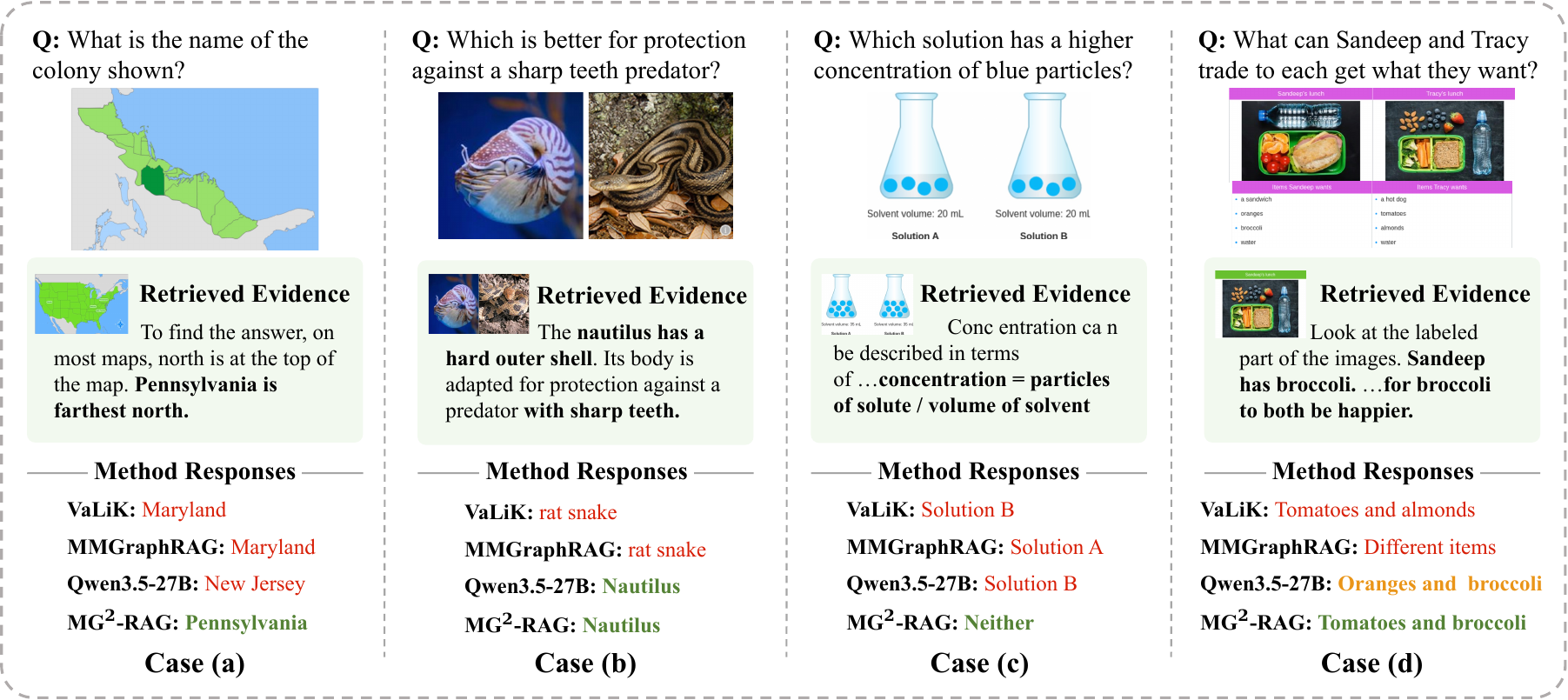}
  \vspace{-0.5em}
  \caption{\textbf{Case studies on Multimodal Reasoning.} MG$^2$-RAG is compared with a baseline MLLM (Qwen3.5-27B) and graph-based approaches (VaLiK and MMGraphRAG). The \textbf{Retrieved Evidence} illustrates the multimodal evidence retrieved by MG$^2$-RAG for answer generation.}
  \label{fig:scienceqa-casestudy}
  \vspace{-0.5em}
\end{figure*}

\paragraph{Multimodal Reasoning Task}
Figure \ref{fig:scienceqa-casestudy} presents additional qualitative examples on \textbf{multimodal reasoning}, where models must jointly interpret visual and textual evidence. Baseline methods frequently produce incorrect or hallucinated answers due to weak cross-modal alignment and limited reasoning capabilities.

MG$^2$-RAG addresses these challenges through two complementary mechanisms.
First, \textbf{modality-preserving multimodal node fusion} tightly aligns fine-grained visual cues with domain-specific textual knowledge. 
For example, in \textbf{Case (a)}, the model links the highlighted region with the retrieved evidence ``Pennsylvania is farthest north'' to correctly identify the colony, while baseline models hallucinate locations such as ``Maryland'' or ``New Jersey.'' 
Likewise, in \textbf{Case (b)}, it associates the animal's visual characteristics with the retrieved description ``hard outer shell,'' correctly distinguishing the Nautilus from visually similar alternatives.

Second, \textbf{multi-granularity graph retrieval} enables structured multi-hop reasoning beyond simple fact matching. 
In \textbf{Case (c)}, MG$^2$-RAG retrieves the governing scientific principle (concentration = particles/volume) and combines it with the visual evidence to infer the correct answer (``Neither''), whereas baseline models are misled by the similar appearance of the flasks. 
Similarly, in \textbf{Case (d)}, graph propagation integrates multiple pieces of textual and visual evidence to correctly determine the traded items (``Tomatoes and broccoli''). 
These examples demonstrate that the synergy between modality-preserving node fusion and graph-based reasoning enables MG$^2$-RAG to solve complex scientific reasoning tasks more accurately and reliably.

\begin{figure*}[!t]
  \centering
  \includegraphics[width=0.99\textwidth]{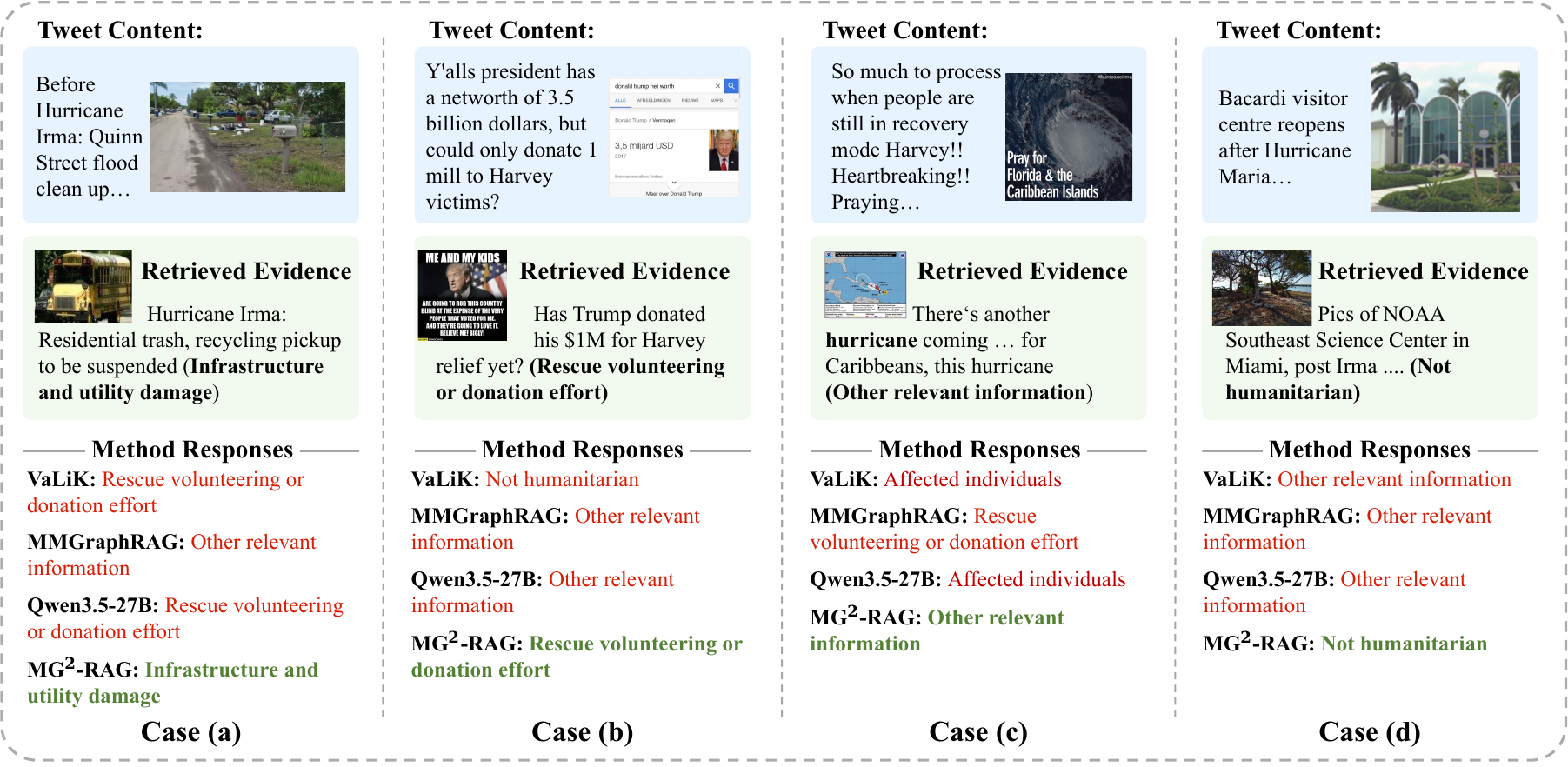}
  \vspace{-0.5em}
  \caption{\textbf{Case studies on Multimodal Classification.} MG$^2$-RAG is compared with a baseline MLLM (Qwen3.5-27B) and graph-based approaches (VaLiK and MMGraphRAG). The \textbf{Retrieved Evidence} illustrates the multimodal evidence retrieved by MG$^2$-RAG for answer generation.}
  \label{fig:crisismmd-casestudy}
  \vspace{-0.5em}
\end{figure*}

\paragraph{Multimodal Classification Task}
Figure \ref{fig:crisismmd-casestudy} further demonstrates the robustness of MG$^2$-RAG on \textbf{multimodal classification} under noisy real-world conditions such as social media posts, where visual and textual signals are often ambiguous, inconsistent, or emotionally charged. 
As a result, baseline models frequently produce generic or incorrect predictions.

MG$^2$-RAG overcomes these challenges through two complementary mechanisms. 
First, \textbf{modality-preserving multimodal node fusion} extracts precise semantic cues by jointly modeling visual and textual evidence. 
For example, in \textbf{Case (a)}, the framework combines street debris with the textual reference ``clean up'' to correctly identify Infrastructure and utility damage. 
Likewise, in Case (b), it aligns donation-related text with the accompanying image to accurately classify the post as ``Rescue volunteering or donation effort.''

Second, \textbf{multi-granularity graph retrieval} filters misleading contextual signals through structured reasoning. 
In \textbf{Case (c)}, instead of being distracted by emotionally charged expressions such as ``Heartbreaking!! Praying...,'' MG$^2$-RAG retrieves factual meteorological updates and correctly classifies the post as ``Other relevant information.'' 
Similarly, in \textbf{Case (d)}, although the keyword ``Hurricane'' biases baseline models toward disaster-related categories, MG$^2$-RAG retrieves evidence about a building reopening and correctly predicts ``Not humanitarian.'' 
These examples demonstrate that the synergy between modality-preserving node fusion and graph-based retrieval enables robust and explainable reasoning in challenging multimodal classification scenarios.

\begin{figure}[!t]
  \centering
  \includegraphics[width=0.99\columnwidth]{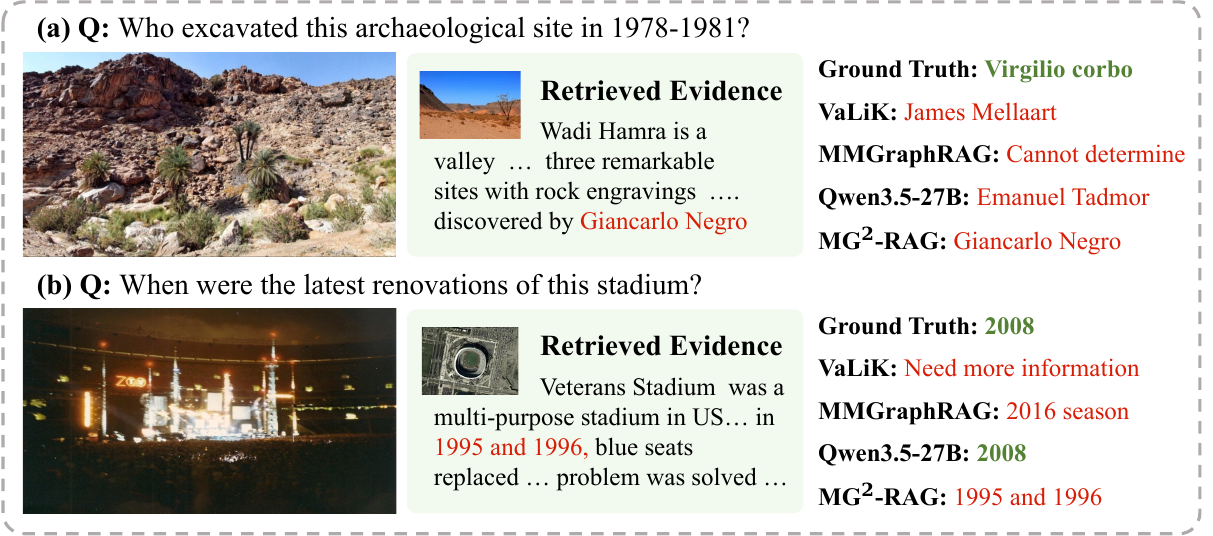}
  \vspace{-0.5em}
  \caption{\textbf{Failure cases of MG$^2$-RAG.} When visual queries contain vague references and lack strong structural cues, retrieval may select plausible but incorrect evidence.}
  \label{fig:casestudy-failure}
  \vspace{-0.5em}
\end{figure}

\paragraph{Failure Cases}
Figure~\ref{fig:casestudy-failure} presents representative failure cases of MG$^2$-RAG.
The framework primarily struggles when queries contain ambiguous visual references (e.g., ``this site'' or ``this stadium'') and the accompanying images provide insufficient discriminative cues.
Under these conditions, entity grounding becomes unreliable, weakening graph propagation and causing retrieval to rely more heavily on dense similarity matching.

Consequently, the retrieved evidence may be semantically related but factually incorrect. 
For example, the model retrieves evidence about \textbf{Giancarlo Negro} instead of the correct archaeologist \textbf{Virgilio Corbo}, and retrieves historical stadium renovation records from \textbf{1995} and \textbf{1996} rather than the target year \textbf{2008}.
These cases highlight a remaining limitation of MG$^2$-RAG: resolving visually ambiguous references requires stronger visual grounding and more effective entity disambiguation during retrieval.

\end{document}